\documentclass[preprint,12pt]{elsarticle}

\usepackage{amssymb}
\usepackage{epsfig, graphicx, lscape, amsmath, amsbsy, amstext}
\usepackage{multirow}

\newcommand{\ve}[1]{{\mbox{\boldmath ${#1}$}}}

\journal{Neurocomputing}

\begin{document}

\begin{frontmatter}



\title{Partial Functional Linear Quantile Regression for Neuroimaging Data Analysis}

\author{Dengdeng Yu}
%%\address[label1]{}

\author{Linglong Kong\corref{ca}}
\cortext[ca]{Corresponding author}
\ead{lkong@ualberta.ca}

\author[]{Ivan Mizera\corref{}}
 
\address{Department of Mathematical and Statistical Sciences, University of Alberta, Edmonton, AB T6G 2G1, Canada}

\author[]{the Alzheimer's Disease Neuroimaging Initiative\fnref{label1}}

\begin{abstract}
We propose a prediction procedure for the functional linear quantile
regression model by using partial quantile covariance techniques and
develop a simple partial quantile
regression (SIMPQR) algorithm to efficiently extract partial quantile
regression (PQR) basis for estimating functional coefficients. We
further extend our partial quantile covariance techniques to
functional composite quantile regression (CQR) defining partial
composite quantile covariance. There are three major contributions.
(1) We define partial quantile covariance between two scalar variables
through linear quantile regression. We compute PQR basis by
sequentially maximizing the partial quantile covariance between the
response and projections of functional covariates. (2) In order to
efficiently extract PQR basis, we develop a SIMPQR algorithm analogous to simple partial least
squares (SIMPLS). (3) Under the homoscedasticity assumption, we extend
our techniques to partial composite quantile covariance and use it to
find the partial composite quantile regression (PCQR) basis. The
SIMPQR algorithm is then modified to obtain the SIMPCQR algorithm. Two
simulation studies show the superiority of our proposed methods. Two
real data from ADHD-200 sample and ADNI are analyzed using our
proposed methods.

\end{abstract}

\begin{keyword}
 Functional linear quantile regression; Partial quantile covariance; PQR basis; SIMPQR; Partial composite quantile covariance; CPQR basis; ADHD; ADNI 

\end{keyword}
 \fntext[label1]{Part of the data used in preparation of this article were obtained from the Alzheimer's Disease Neuroimaging
Initiative (ADNI) database (adni.loni.ucla.edu). As such, the investigators within the ADNI contributed to
the design and implementation of ADNI and/or provided data but did not participate in analysis or writing
of this report. A complete listing of ADNI investigators can be found at: {\sl http://adni.loni.ucla.edu/wp-content/
 uploads/how\_to\_apply/ADNI\_Acknowledgement\_List.pdf}}
\end{frontmatter}


\section{Introduction}

Nowadays, there is great need in the analysis of complex neuroimaging
data obtained from various cross-sectional and clustered neuroimaging
studies. These neuroimaging studies are essential to advancing our
understanding of the neural development of neuropsychiatric and
neurodegenerative disorders, substance use disorders, the normal brain
and the interactive effects of environmental and genetic factors on
brain structure and function. Such large imaging studies include the
ADNI (Alzheimer's Disease Neuroimaging Initiative), the longitudinal
magnetic resonance imaging (MRI) study of schizophrenia, autism, and
attention deficit hyperactivity disorder (ADHD), the NIH human
connectome project, among many others. Neuroimaging studies usually
collect structural, neurochemical, and functional images over both
time and space
\cite{fass2008imaging,friston2009modalities,niedermeyer2005electroencephalography}.
These structural, neurochemical, and functional imaging modalities
include computed axial tomography (CT), diffusion tensor imaging
(DTI), functional magnetic resonance imaging (fMRI), magnetic
resonance imaging (MRI), magnetic resonance spectroscopy (MRS),
positron emission tomography (PET), single photon emission tomography
(SPECT), electroencephalography (EEG), and magnetoencephalography
(MEG), among many others. For instance, by using anatomical MRI,
various measures of the morphology of the cortical and subcortical
structures (e.g., hippocampus) are extracted to understand
neuroanatomical differences in brain structure across different
populations \cite{domschke2010imaging,scharinger2010imaging}. In DTI,
various diffusion properties and fiber tracts are extracted for
quantitative assessment of anatomical connectivity across different
populations
\cite{basser1994mr,zhu2011fadtts,zhu2012multivariate,zhu2007statistical}.
Functional images, such as resting-state functional MRI (rsfMRI), have
been widely used in behavioral and cognitive neuroscience to
understand functional segregation and integration of different brain
regions across different populations
\cite{huettel2004functional,penny2011statistical}.

\begin{figure}[!ht]
\includegraphics[width=.50\textwidth]{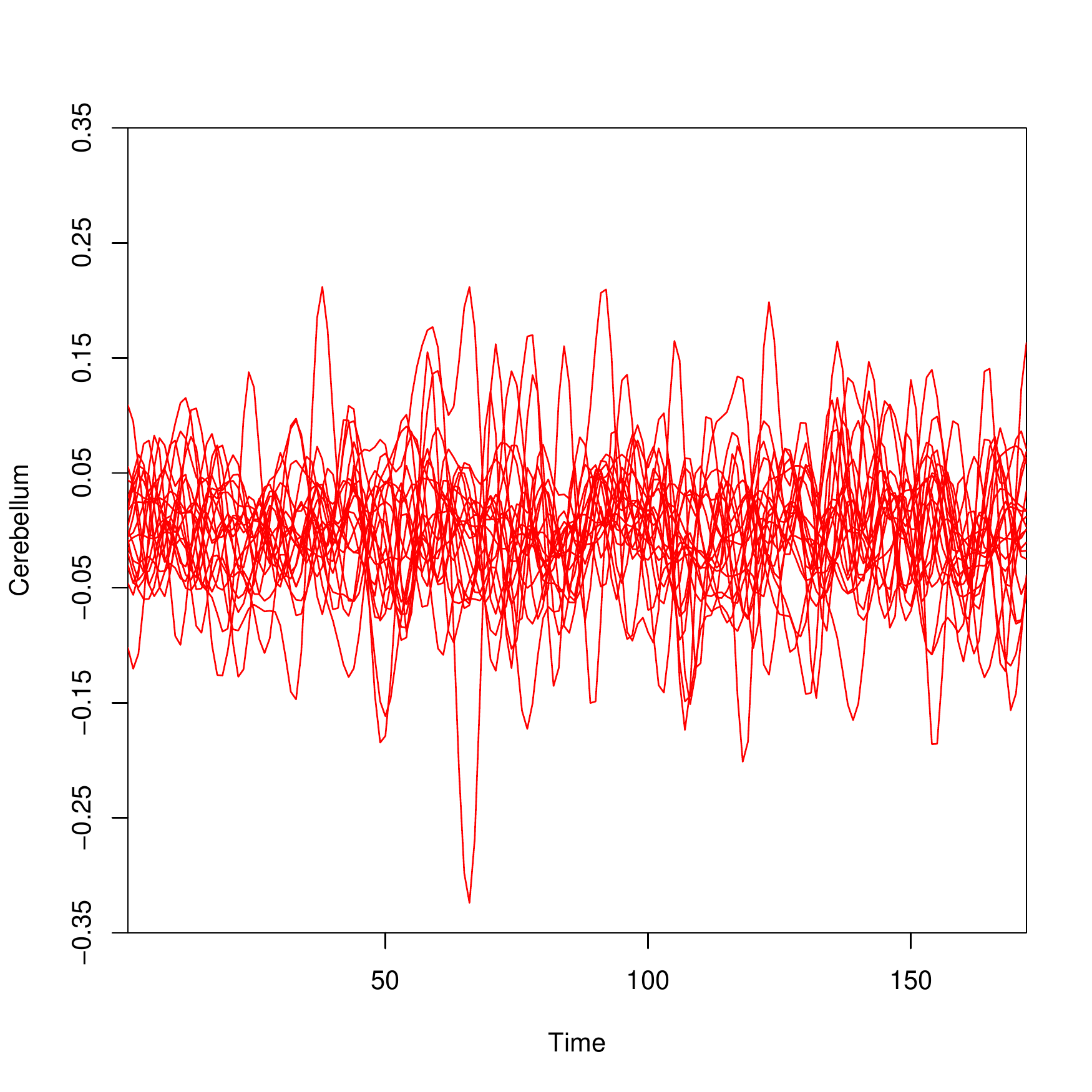}%
\includegraphics[width=.50\textwidth]{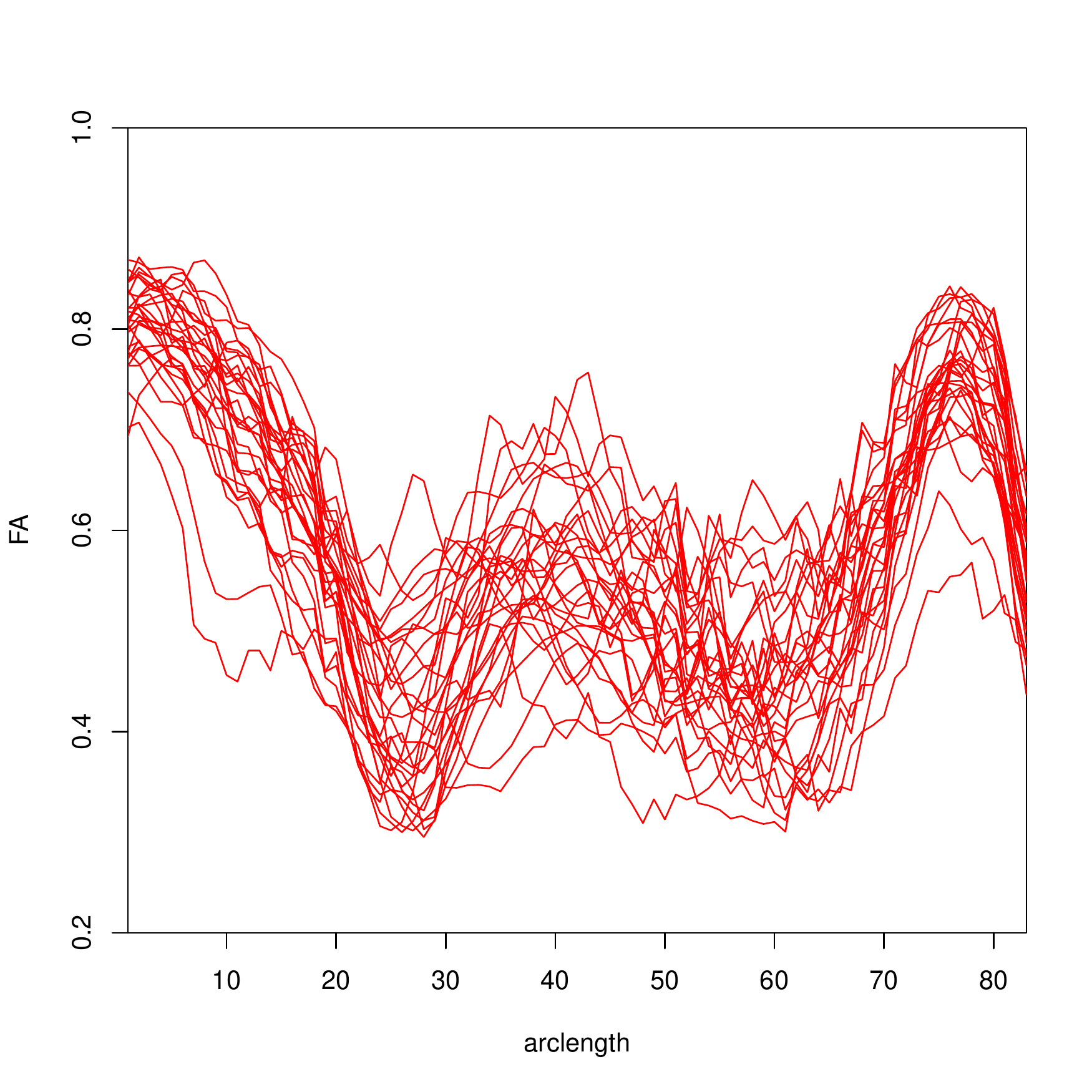}
\caption{ Representative functional neuroimaging data: (Left) the estimated hemodynamic response functions (HRF) corresponding to resting-state from 20 children at NYU  from the ADHD-200 Sample Initiative Project and (Right) fractional anisotropy (FA)  along the midsagittal corpus callosum (CC) skeleton from 30 randomly selected subjects from the NIH Alzheimer's Disease Neuroimaging Initiative (ADNI) study.} \label{nc_fig1}
\end{figure}

A common feature  of many  imaging techniques is that   massive
functional data are observed/calculated at the same design points, such as time for functional images (e.g., PET and fMRI) and arclength for structure imaging (e.g. DTI).
As an illustration, we present two smoothed functional data that we encounter in neuroimaging studies.
First,  we consider the BOLD rsfMRI signal, which  is based on hemodynamic responses secondary to resting-state. 
We plot the estimated hemodynamic response functions (HRF) with 172 time courses from 20 randomly selected children  at a selected region of interest (ROI) of 
Anatomical Automatic Labeling (AAL) atlas \cite{tzourio2002automated} from the New York University (NYU) Child Study Center from the ADHD-200 Sample Initiative Project.   Although the canonical form of the HRF   is often used,    when applying rsfMRI in a clinical
population with possibly altered hemodynamic responses (Figure \ref{nc_fig1} (a)), using the subject's own HRF in
rsfMRI data analysis may be advantageous because HRF variability is greater across subjects
than across brain regions within a subject \cite{aguirre1998variability, lindquist2009modeling}.  We
 are particularly interested in delineating the structure of the
variability of the HRF and their capacity of predicting ADHD index with a set of covariates of interest, such
as diagnostic group  \cite{lindquist2008statistical}. Secondly, we   plot  one diffusion property, called fractional
anisotropy (FA), measured at 83 grid points along  the midsagittal corpus callosum (CC) skeleton
  (Figure \ref{nc_fig1} (b)) from 30 randomly selected
infants from the NIH Alzheimer's Disease Neuroimaging Initiative (ADNI) study. The corpus callosum (CC) is the largest
fiber tract in the human brain and is a topographically organized structure. It is responsible for much of
the communication between the two hemispheres and connects homologous areas in the two
cerebral hemispheres. Scientists are
particularly interested in delineating the structure of the
variability of these functional FA data and their  prediction ability on mini-mental state examination (MMSE) with a set of covariates of interest, such
as genetic information. MMSE is 
one of the most widely used screening tests on Alzheimer's Disease to provide brief and objective
measures of cognitive functioning \cite{tombaugh1992mini}.   We will systematically investigate 
these two prediction problems using functional imaging data over time or space in Section \ref{ns_re} after we develop our methodology.

A functional linear regression model, where the responses such as the
neurological or clinical outcomes (e.g. ADHD index or MMSE) are
modeled by a set of scalar covariates and functional covariates of
interest (e.g. HRF along time courses or FA along arclength), is a
powerful statistical tool for addressing these scientific questions
\cite{goldsmith2011penalizedJCGS,goldsmith2011penalized,zhu2011fadtts,zhu2012multivariate}.
In particular, denoting the neurological or clinical outcome of the
$i$-th subject by $y_i$, $i=1,\dots,n$, the functional linear
regression model is of the form
\begin{equation}\label{pfqreq1}
y_i=\alpha+\ve x_i^T\ve \beta + \int_0^1 \ve z_i^T(t) \ve \gamma(t)dt +\epsilon_i,
\end{equation}
where $\alpha$ is the intercept, $\ve \beta=(\beta_1, \cdots,
\beta_p)^T$ is a $p\times 1$ vector of coefficients, ${\ve
  x}_i=(x_{i1},\cdots,x_{ip})^T$ is a $p\times 1$ vector of scalar
covariates of interest, $\ve \gamma(t)=(\gamma_1(t), \cdots,
\gamma_q(t))^T$ is a $q\times 1$ vector of coefficient functions of
$t$, ${\ve z}_i(t) = (z_{i1}(t),\cdots,z_{iq}(t))^T$ is a $q\times 1$
vector of functional covariates, and $\epsilon_{i}$ is a random error.
It is usually assumed that $\epsilon_i$ is independent and identical
copy of normal distribution with zero mean and variance $\sigma^2$.
For simplicity, we let $t \in [0,1]$. Model ~\eqref{pfqreq1} is a
generalization of the classical linear regression model corresponding
to the case $\ve \gamma(t)$ is a constant. If it is not constant, the
contributions of $ \ve z_i(t)$ characterized by $\ve \gamma(t)$ change
in terms of $t$. The model has been well studied and applied in many
fields including neuroimaging data analysis
\cite{aneiros2006semi,james2009functional,lian2011functional,shin2009partial}.
To facilitate the estimation of $\ve \gamma(t)$, we usually require
that it satisfies certain smoothness conditions and restrict it onto a
functional space. For example, we may require that it its second
derivative exists and that the square of $\gamma(t)$ is integrable,
that is, $\ve \gamma(t) \in L_2[0,1]$. Even in such a case, the
estimation is still an infinite-dimensional problem.

The common practice is to project $\ve \gamma(t)$ into a functional space with a finite functional basis. There are three major methods to  choose
the functional basis: general basis, functional principal component basis (fPC), and partial least square basis (PLS). There are various options on the selection 
of general basis, for example B-spline basis \cite{cai2006prediction,cardot2003spline}, wavelet basis \cite{zhao2012wavelet} and so on. In order to provide a good approximation of 
the functional coefficients, a large number of basis should be chosen. However, this may cause overfitting of the model  and to remedy that various penalty methods have been proposed 
\cite{crambes2009smoothing,zhao2014wavelet}. The fPC method has been extensively studied \cite{hall2007methodology,lee2012sparse} where the fPC of ${\ve z}_i(t)$ 
serve as the basis. Its generalization to the reproducing kernel Hilbert space (RKHS) was proposed by Cai and Yuan \cite{cai2012minimax,yuan2010reproducing} who also studied its minimax rates. Although fPC basis are more data-adapted than the general basis as they use the information of functional covariates and the formed space 
can explain most of the variation of ${\ve z}_i(t)$, it is not necessary all the fPC basis will contribute to the variation of the responses. 
Therefore, another appealing choice is the PLS basis which use both the information of functional covariates and the responses. The PLS basis use the linear projects of ${\ve z}_i(t)$ which best predict the responses \cite{delaigle2012methodology}. 

An alternative to model \eqref{pfqreq1} is the functional linear
quantile regression where the conditional quantiles of the responses
are modeled by a set of scalar covariates and functional covariates.
There are at least three advantages to use conditional quantiles
instead of conditional means. First, quantile regression, in
particular median regression, provides an alternative and complement
to mean regression while being resistant to outliers in responses. It
is more efficient than mean regression when the errors follow a
distribution with heavy tails. Second, quantile regression is capable
of dealing with heteroscedasticity, the situation when variances
depend on some covariates. More importantly, quantile regression can
give a more complete picture on how the responses are affected by
covariates: for example, some tail behaviors of the responses
conditional on covariates. For more background on quantile regression,
see the monograph of Koenker \cite{koenker2005quantile}. In our case,
we consider functional linear quantile regression: for given $\tau \in
(0,1)$,
\begin{equation}\label{pfqreq2}
Q_{\tau}\left(y_i|\ve x_i,\ve z_i(t)\right)=\alpha_\tau+\ve x_i^T\ve \beta_\tau + \int_0^1 \ve z_i^T(t) \ve \gamma_\tau(t)dt,
\end{equation}
where $Q_{\tau}\left(y_i|\ve x_i,\ve z_i(t)\right)$ is the $\tau$-th
conditional quantile of $y_i$ given covariates $\ve x_i$ and ${\ve
  z}_i(t)$, $\alpha_\tau$ is the intercept, $\ve
\beta_\tau=(\beta_{1\tau}, \cdots, \beta_{p\tau})^T$ is a $p\times 1$
vector of coefficients and $\ve \gamma_\tau(t)=(\gamma_{1\tau}(t),
\cdots, \gamma_{q\tau}(t))^T$ is a $q\times 1$ vector of coefficient
functions. In the existing literature, model \eqref{pfqreq2} has been
well studied and various methods have been proposed. As in functional
linear regression, to estimate functional coefficients $\ve
\gamma_\tau(t)$ it is convenient to restrict it in a functional space
with a finite basis. Similarly, general basis like B-spline can be
used to approximate the coefficient functions
\cite{cardot2005quantile,sun2005semiparametric}. fPC basis have also
been throughly investigated with and without scalar covariate $\ve
x_i$ while only one functional covariate presents
\cite{kato2012estimation, lu2014functional,tang2014partial}. However,
there is no analogue to the PLS basis method in functional linear
regression model. Therefore, none of the existing methods for model
\eqref{pfqreq2} is able to provide more efficient prediction by
extracting information from the responses.

In this paper, we propose a prediction procedure for the functional
linear quantile regression model \eqref{pfqreq2} by using partial
quantile covariance techniques and develop an algorithm inspired by
simple partial linear regression, SIMPLS \cite{de1993simpls}, to
efficiently extract partial quantile regression (PQR) basis for
estimating functional coefficients. We further extend our partial
quantile covariance techniques to functional composite quantile
regression (CQR) \cite{zou2008composite} by defining partial composite
quantile covariance. The major contributions of this paper can be
summarized as follows. We first define partial quantile covariance
between two scalar variables through linear quantile regression.
Motivated by extracting PLS basis in functional linear regression, we
found PQR basis by sequentially maximizing the partial quantile
covariance between the response and projections of functional
covariates. In order to efficiently extract PQR basis, we develop a
simple partial quantile regression (SIMPQR) algorithm analogue to
SIMPLS. Under the homoscedasticity assumption, we extend our
techniques to partial composite quantile covariance and use it to find
the partial composite quantile regression (PCQR) basis. The SIMPQR
algorithm is then modified to obtain the SIMPCQR algorithm.

The rest of this paper is organized as follows. In Section 2, we
define partial quantile covariance and describe how to use it to extract PQR basis in functional linear quantile regression model. 
In Section 3, we develop the SIMPQR algorithm and discuss its properties. We discuss how to calculate the PCQR basis by using 
partial composite quantile covariance and propose the SIMPCQR algorithm in Section 4. 
Two sets of   simulation studies are presented in Section 5 with the known ground truth to
examine the finite sample performance of our proposed methodology. In Section 6, we use PQR and PCQR to predict ADHD index and 
MMSE using data from NYU site from ADHD-200 sample and ADNI, respectively. Some discussions and future research directions are given in Section 7. 

\section{Partial Functional Linear Quantile Regression}
In model \eqref{pfqreq2}, we assume without loss of generality that $t
\in [0,1]$ and restrict the functional coefficients $\ve\gamma_\tau(t)
\in L_2[0,1]$. 
For simplicity, we assume $q=1$, that is, we only
consider one functional covariate. The extension of our methodology to
more functional covariates is straightforward. 
The estimation
$\ve\gamma_\tau(t)$ is in general a difficult question as it lies in
an infinite-dimensional space. However, if it can be well approximated
in a finite element space, say, $H[0,1]$, the solution for the model
\eqref{pfqreq2} can be found. Let $b_{k\tau}(t)$, $k=1,\dots,K$ be a
basis of $H[0,1]$ and $\ve\gamma_\tau(t) =
\sum_{k=1}^K\ve\gamma_{k\tau}b_{k\tau}(t)$. Model \eqref{pfqreq2}
can be then rewritten as
\begin{equation}\label{pfqreq3}
Q_{\tau}\left(y_i|\ve x_i,\ve z_i(t)\right)=\alpha_\tau+\ve x_i^T\ve \beta_\tau + \sum_{k=1}^K \ve z_{ki} \ve \gamma_{k\tau},
\end{equation}
where $\ve z_{ki}= \int_{0}^1\ve z_i(t)b_{k\tau}(t) dt$. Model
\eqref{pfqreq3} is simply a linear quantile regression problem, which
is essentially a linear programing problem; its solutions can be
obtained by many algorithms---for example, the simplex method
\cite{boyd2004convex}, the interior point method
\cite{koenker2005quantile}, the MM algorithm \cite{hunter2000quantile}
and many others, already implemented in various statistical softwares
like \texttt{quantreg} in \texttt{R} \cite{quantreg2013}. 

In the literature, there are many methods devoted to find the crucial
basis functions in model \eqref{pfqreq3}. The general basis B-spline
was proposed and studied by Cardot and others
\cite{cardot2005quantile,sun2005semiparametric}. In various models,
fPC basis has also been studied \cite{kato2012estimation,
  lu2014functional,tang2014partial}. However, neither basis does use
information of the responses and hence they are less efficient to do
prediction. In this session, as the motivation of our proposal, we
first review the PLS basis in model \eqref{pfqreq1} where both
information of the functional covariates and the responses are used to
choose the basis functions. Then we propose our developed methodology
to choose basis for model \eqref{pfqreq2}, namely, partial quantile
regression (PQR) basis.

In functional linear regression model \eqref{pfqreq1}, the first PLS basis is chosen to be 
\begin{equation}\label{pfqreq4}
b(t) = \arg_{b(t)}\min_{\alpha,\beta,b(t)}\sum_{i=1}^n\left(y_i-\alpha-\ve x_i^T\ve \beta - \int_0^1 \ve z_i(t)b(t)dt\right)^2, 
\end{equation}
which is the analogue to the partial least square regression in
multivariate analysis. The subsequent basis is chosen by iteratively
using \eqref{pfqreq4} after taking account of and subtracting the
information from previous basis. For more details, see Delaigle and
Hall \cite{delaigle2012methodology}. The essential idea of criteria
\eqref{pfqreq4} is to find a direction $b(t)$ so that the projection
of $\ve Z(t)$ on it explains as much as possible the variation of the
response after adjusting some covariates. Therefore, as shown in
\cite{delaigle2012methodology}, it is equivalent to find a basis
$b(t)$ such that the partial covariance
\begin{equation}\label{pfqreq5}
COV\left(Y-\alpha-\ve X\ve \beta, \int_0^1 \ve Z(t)b(t)dt\right) 
\end{equation}
is maximized, where $Y = (y_1,\dots,y_n)^T$, $\ve X = (\ve x_1,\dots,\ve x_n)^T$ and $\ve Z(t) = (\ve z_1(t),\dots,\ve z_n(t))^T$. Based on this equation, Delaigle and Hall \cite{delaigle2012methodology} found an equivalent space with the same dimension as 
the PLS space and proved the estimation and precision consistency. 

The parameters in model \eqref{pfqreq2} are estimated by solving
\begin{equation}\label{pfqreq6}
\min_{\alpha,\beta,b(t)}\sum_{i=1}^n\rho_\tau\left(y_i-\alpha-\ve x_i^T\ve \beta - \int_0^1 \ve z_i (t)b(t)dt\right), 
\end{equation}
where $\rho_\tau(u) =u(\tau-I(u<0))$ is the quantile loss function
\cite{koenker2005quantile} with $I$ as the indicator function. When
$\tau=0.5$, the loss is $\rho_\tau(u) =|u|/2$ and the results is then
the median, or least absolute deviation (LAD) regression. To adapt to
the idea of PLS basis, that is, to find a direction $b_\tau(t)$ so
that the projection of $\ve z(t)$ on it contributes as much as
possible to predict the quantile of the response after adjusting some
covariates, we first propose the concepts of quantile covariance (QC)
and partial quantile covariance (PQC). For given $\tau \in (0,1)$ and
a random variable $X$, the partial quantile covariance $COV_{qr}(Y,Z)
$ between two random variables $Y$ and $Z$ is of the form
\begin{equation}\label{pfqreq7}
COV_{qr}(Y,Z) =\arg_{\gamma_\tau}\inf_{\alpha,\beta_\tau,\gamma_\tau}E\left(\rho_\tau\left(Y-\alpha-\beta_\tau X-\gamma_\tau Z \right)\right), 
\end{equation}
where we first normalize $Z$ to have mean zero and variance one. If
there is no $X$, then $COV_{qr}(Y,Z) $ is quantile covariance between
$Y$ and $Z$. The quantile covariance measures the contribution of $Z$
to the $\tau$-th quantile of $Y$. It was first proposed and studied by
Dodge and Whittaker \cite{dodge2009partial} in the context of partial
quantile regression. Li, Li and Tsai \cite{li2014quantile} proposed a
similar concept of quantile correlation and used it to study quantile
autoregressive model.

To find the partial quantile regression basis (PQR), similar to that of PLS to maximize the covariance we propose to compute the $b_\tau(t)$ by maximizing 
\begin{equation}\label{pfqreq8}
COV_{qr}\left(Y-\alpha_\tau-\ve X\ve \beta_\tau, \int_0^1 \ve Z(t)b_\tau(t)dt\right). 
\end{equation}
The subsequent basis is computed by iteratively maximizing
\eqref{pfqreq8} after taking account of and subtracting the
information of the previous basis. Let $\ve Z_k = \int_0^1\ve
Z(t)b_{k\tau}(t)dt$, where $b_{k\tau}(t)$ is the $k$-th PQR basis.
Denote $\ve Z_{(k+1)}(t)$ as $\ve Z(t)$ after subtracting the
information from the first $k$ basis. Then the $(k+1)$-th basis
$b_{(k+1)\tau}(t)$ is obtained by maximizing the partial quantile
covariance
\begin{equation}\label{pfqreq9}
COV_{qr}\left(Y-\alpha_\tau-\ve X\ve \beta_\tau-\sum_{j=1}^k\ve Z_{j} \ve \gamma_{j\tau}, \int_0^1 \ve Z_{(k+1)}(t)b_\tau(t)dt\right). 
\end{equation}
We will discuss detailed algorithms in the next section. Once we find
an adequate number, $K$ of functional basis elements, we have the
approximation model \eqref{pfqreq3}, where the parameters are obtained
by minimizing
\begin{equation}\label{pfqreq10}
\sum_{i=1}^n\rho_\tau\left(y_i-\alpha_\tau-\ve x_i^T\ve \beta_\tau - \sum_{k=1}^K \ve z_{ki} \ve \gamma_{k\tau}\right).
\end{equation}
The number of PQR basis can be chosen use BIC or cross validation (CV) as in choosing the number of fPC basis adapted by Kato and other authors 
\cite{kato2012estimation,tang2014partial,lu2014functional}.  

\section{SIMPQR Algorithm}
In this section, we propose a simple partial quantile regression
(SIMPQR) algorithm to iteratively extract the PQR basis from the
functional covariates $\ve z(t)$. Similar algorithm has been studied
by Dodge and Whittaker \cite{dodge2009partial} in partial quantile
regression with multiple covariates. It is parallel to the SIMPLS for
partial least square regression \cite{de1993simpls}. The motivation is
to subsequently maximize \eqref{pfqreq8} after accounting and
subtracting the information of the previous basis. To simplify the
description of the SIMPQR algorithm, we will drop the scalar
covariates $\ve x$ in model \eqref{pfqreq9} in this section. Let
$0<t_1<\dots<t_m<1$ denote the discretized sample points for the
functional covariates and we assume they are equally spaced. Recall
that we set $q=1$ and we focus on only one functional covariate $\ve
z(t)$. The SIMPQR algorithm is described as follows.
\begin{description}
\item {\it Step 1, Initialization: } Normalize $\ve z_i(t_j)$ for each $j$ so that it has mean zero and variance one. 
 \item {\it Step 2, Repeat: }
 \begin{enumerate}
\item Compute a functional basis $b_\tau(t_j) = COV_{qr}(Y,\ve Z(t_j))$ for $j=1,\dots,m$ and rescale it to have $\sum b^2_\tau(t_j)=1$. 
\item Project $\ve z_i(t_j)$ onto the basis $(b_\tau(t_1),\dots,b_\tau(t_m))^T$ to obtain $\ve z_{i} = \sum \ve z_i(t_j) b_\tau(t_j)$. Denote $\ve Z = (\ve z_1,\dots,\ve z_n)^T$ as the projections for each subjects. 
\item Predict $\ve z_i(j)$ by using simple linear regression with the projection $\ve z_i$ as the covariate, denoting the result by $\hat{\ve z}_i(j)$. 
\item Subtract the information from the projection $\ve z_i$ by replacing $\ve z_i(j)$ by their residuals 
$\ve z_i(j)-\hat{\ve z}_i(j)$. 
\end{enumerate}
\item {\it Step 3, Stop: }  Check stopping criterion and retain the projections $\ve Z_1,\dots,\ve Z_K$. 
\item {\it Step 4, Model: }  Fit the model \eqref{pfqreq3} by minimizing equation \eqref{pfqreq10}.
 \end{description}

The SIMPQR algorithm follows the same line of SIMPLS with the
covariance being replaced by quantile covariance. The nature of our
proposed quantile covariance implies that it is not necessary to
adjust the response~$Y$ each time after a new basis is obtained. The
resulting functional basis is orthogonal to each other due to the
prediction in step 2.3. However, it is worth noting that due to the
nonadditivity of conditional quantiles, we need to fit model
\eqref{pfqreq3} after all the basis elements are picked out, instead
of estimating the coefficients of each basis projections once they are
chosen.

\section{Partial Functional Linear Composite Quantile Regression}
Despite the success of quantile regression (QR), its relative efficiency to the least square regression can be arbitrarily small \cite{koenker2005quantile,zou2008composite}. 
Composite quantile regression (CQR) proposed by Zou and Yuan \cite{zou2008composite} inherits some good properties of QR and is capable of providing 
more efficient estimators under certain conditions.  Given two random variables $X$ and $Y$ and quantile level set $0<\tau_1<\dots<\tau_L<1$, the CQR parameters $(\alpha_{\tau_1},\dots,\alpha_{\tau_L},\beta)$ are defined as 
 \begin{equation}\label{pfqreq11}
\inf_{\alpha_{\tau_1},\dots,\alpha_{\tau_L},\beta}E\sum_{l=1}^L\left(\rho_{\tau_l}\left(Y-\alpha_{\tau_l}-\beta X \right)\right), 
\end{equation} 
where $\rho_{\tau_l}$ is the $\tau_l$-th quantile loss function. Under the homoscedasticity assumption, that is, the model errors do not depend on 
covariates, all conditional regression quantiles are parallel and they have the same slope $\beta$ but different intercepts. The CQR is equivalent to 
fit QR at different quantile levels. However, CQR estimators are more efficient. 

For given $0<\tau_1<\dots<\tau_L<1$ and a random variable $X$ similar
to \eqref{pfqreq7}, the partial composite quantile covariance (PCQC)
$COV_{cqr}(Y,Z) $ between two random variable $Y$ and $Z$ is of the
form
\begin{equation}\label{pfqreq12}
COV_{cqr}(Y,Z) =\arg_{\gamma}\inf_{\alpha_{\tau_1},\dots,\alpha_{\tau_L},\beta,\gamma}E\sum_{l=1}^L\left(\rho_{\tau_l}\left(Y-\alpha_{\tau_l}-\beta X-\gamma Z \right)\right), 
\end{equation}
where we first normalize $Z$ to have mean zero and variance one. If
there is no $X$, then $COV_{cqr}(Y,Z) $ is composite quantile
covariance (CQC) between $Y$ and $Z$. The composite quantile
covariance measures the contribution of $Z$ to the quantiles of $Y$ at
levels $0<\tau_1<\dots<\tau_L<1$. There are some connections between
composite quantile covariance and covariance; however, these are beyond
the scope of this paper and we plan to discuss them elsewhere.

With the definition of PCQC, we can obtain the PCQR basis for functional linear composite quantile regression by maximizing 
\begin{equation}\label{pfqreq13}
COV_{cqr}\left(Y-\ve X\ve \beta, \int_0^1 \ve Z(t)b(t)dt\right),
\end{equation}
for a given quantile level set $0<\tau_1<\dots<\tau_L<1$. The
subsequent basis is computed by iteratively maximizing
\eqref{pfqreq13} after accounting and subtracting the information of
the previous basis. Once the PCQR basis is found, the functional
linear composite quantile regression can be easily fitted by a linear
program, for example \texttt{quantreg} in \texttt{R}
\cite{quantreg2013}. The algorithm to compute the PCQR basis follows
the same line of SIMPQR in the last section; we only need to replace
$COV_{qr}$ by $COV_{cqr}$ in step 2.1 and keep the rest unchanged.

\section{Simulation Studies}
In this section, we investigate the finite sample performance of our
proposed prediction methods, namely partial quantile regression (PQR)
basis and partial composite quantile regression (PCQR) basis methods.
We compare them with the fPC basis method in functional linear
quantile regression (QRfPC) and functional linear composite quantile
regression (CQRfPC) models. In addition, we compare them with PLS
basis and fPC basis methods in functional linear regression model. We
conduct our simulations in two settings where the first one is in
favor of the fPC basis and the second one is a more general case. Both
simulations show superior or comparable performance of our proposed
methods.

{\sl Simulation I.} In this simulation, we adapt the setup in Kato \cite{kato2012estimation}.  In particular, the model is of the form 
\begin{eqnarray*}
Y &=& \int_{0}^{1} \gamma(t) Z(t) dt + \varepsilon,\\
\gamma(t)&=& \sum_{j=1}^{50} \gamma_j \phi_j(t); ~~\gamma_1=0.5, \gamma_j=\frac{20}{3}(-1)^{j+1} j^{-2}, j \geq 2, \phi_j(t)=2^{1/2}\cos(j \pi t),\\
Z(t) &=& \sum_{j=1}^{50} v_j U_j \phi_j(t);~~ v_j = (-1)^{j+1} j^{-1.1/2},U_j \sim U[-3^{1/2},3^{1/2}].
\end{eqnarray*}
Each $X_i(t)$ was observed at $m=201$ equally spaced grid points on $[0,1]$. We choose the sample size $n$ to be $100$, $200$, and $500$. The error 
$\varepsilon$ follows either Gaussian with mean zero and variance one or Cauchy distribution.
In this design we have
\begin{equation*}
Q_\tau(Y|X) = F_{\varepsilon}^{-1}(\tau) + \int_0^1 \gamma(t) Z(t) dt,
\end{equation*}
where $F_{\varepsilon}$ is the cumulative distribution function of $\varepsilon$. It should be pointed out that 
the simulation set up is in favor of fPC basis methods as the functional coefficients lie on the same fPC space of 
functional covariates. It is expected that fPC basis methods may be superior to other methods. 

To facilitate the comparison, we set $\tau=0.5T$ for QR methods and $\tau_l=l/(1+L)$ with $L=9$ for CQR methods. One criteria we use is 
the mean integrated errors (MISE) of the functional coefficients,  
\begin{eqnarray*}
\mbox{MISE} =\frac{1}{S}\sum_{s=1}^S
\sum_{j=1}^m\left(\hat\gamma_s(t_j)-\gamma(t_j)\right)^2 
= \mbox{Bias}^2+\mbox{Var},
\end{eqnarray*}
where 
\begin{equation*}
\mbox{Bias}^2 =
\sum_j\left(\frac{1}{S}\sum_s\hat\gamma_s(t_j)-\gamma(t_j)\right)^2
\end{equation*}
and 
\begin{equation*}
\mbox{Var}=\frac{1}{S}\sum_s
\sum_j\left(\hat\gamma_s(t_j)-\frac{1}{S}\sum_s\hat\gamma_s(t_j)\right)^2.
\end{equation*}

\begin{figure}[htbp]
 \centering
  \begin{minipage}[t]{.49\textwidth}
     \includegraphics[scale=0.36]{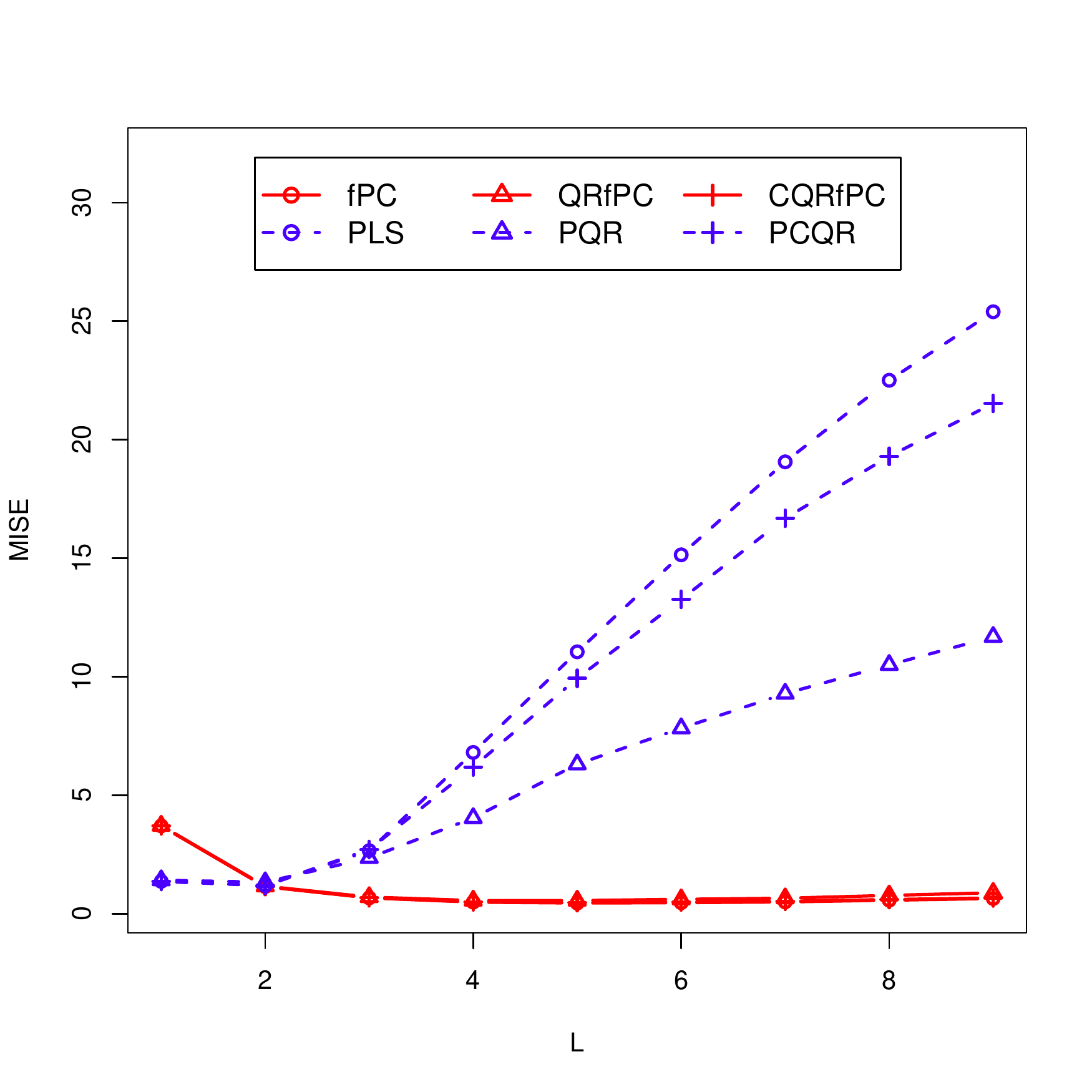}
 \end{minipage}
  \begin{minipage}[t]{.49\textwidth}
     \includegraphics[scale=0.385]{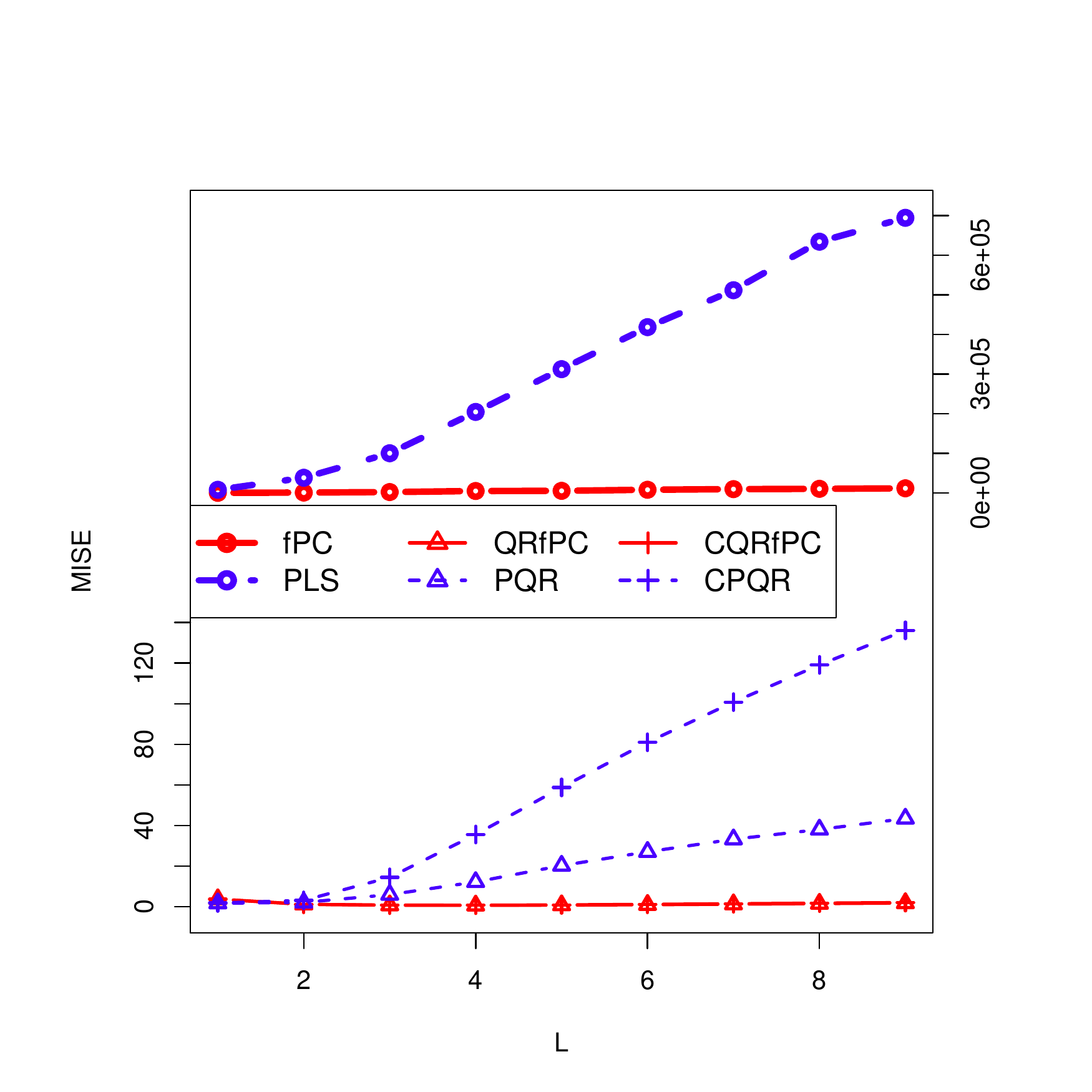}
 \end{minipage}
   \begin{minipage}[t]{.49\textwidth}
     \includegraphics[scale=0.36]{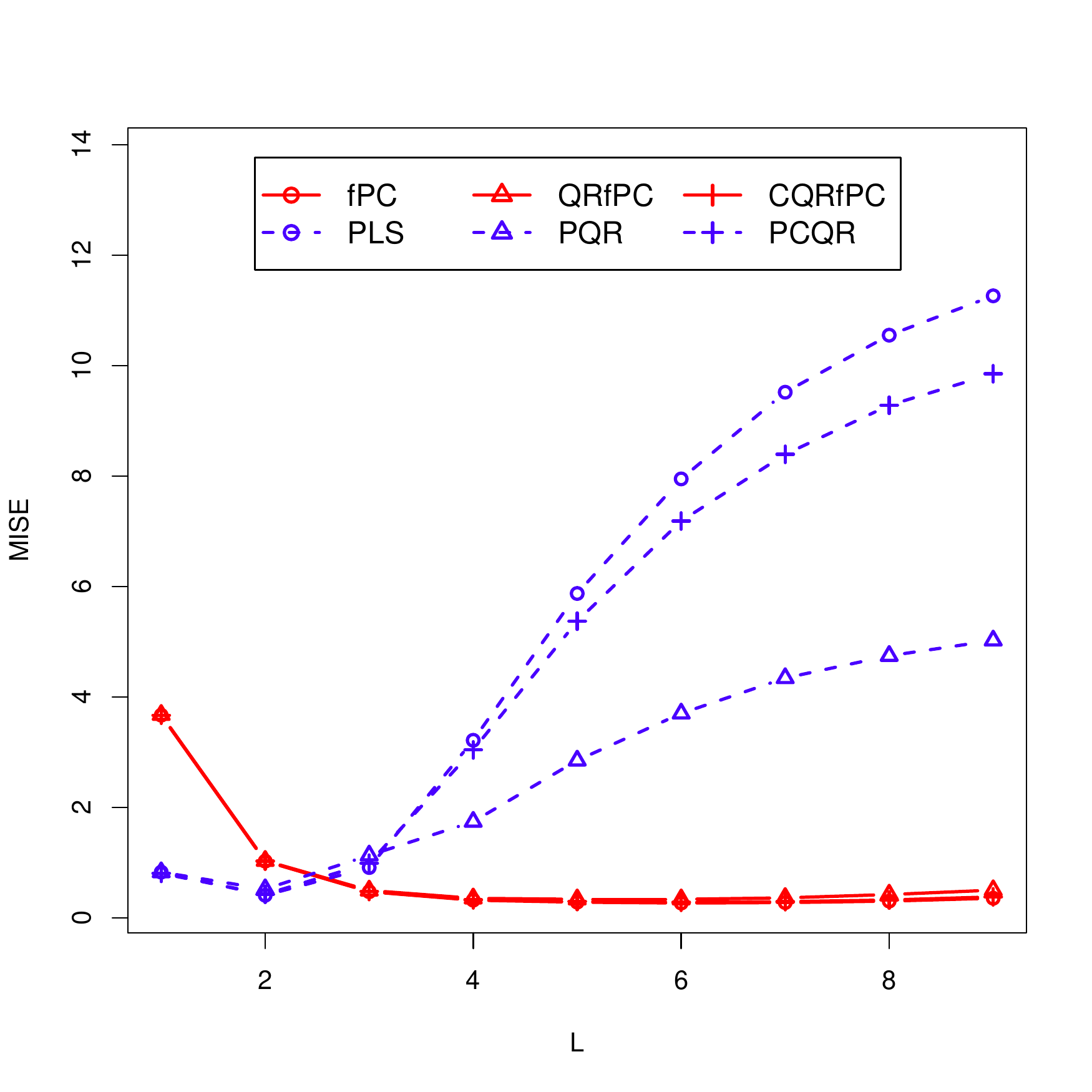}
 \end{minipage}
   \begin{minipage}[t]{.49\textwidth}
     \includegraphics[scale=0.385]{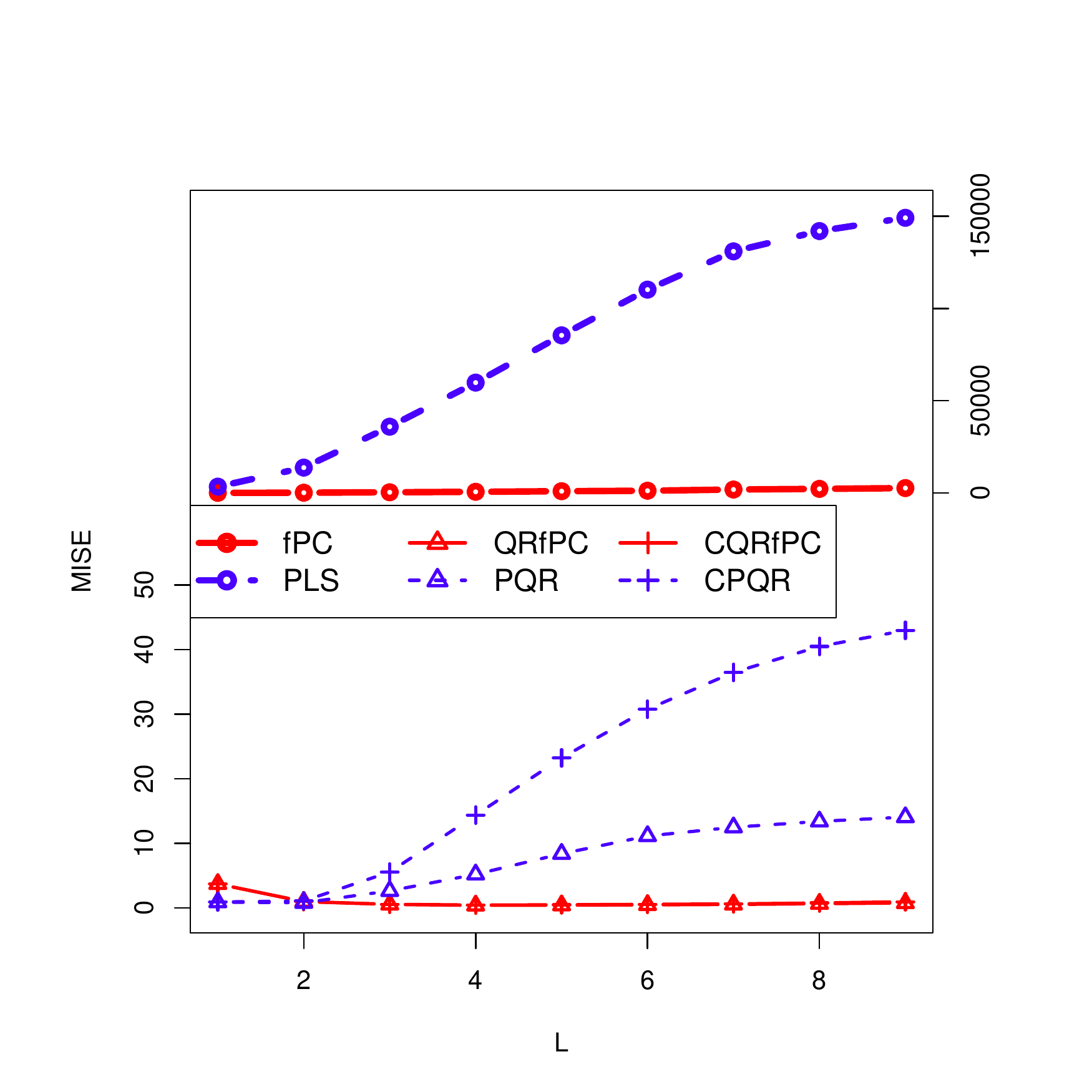}
 \end{minipage}
  \begin{minipage}[t]{.49\textwidth}
     \includegraphics[scale=0.36]{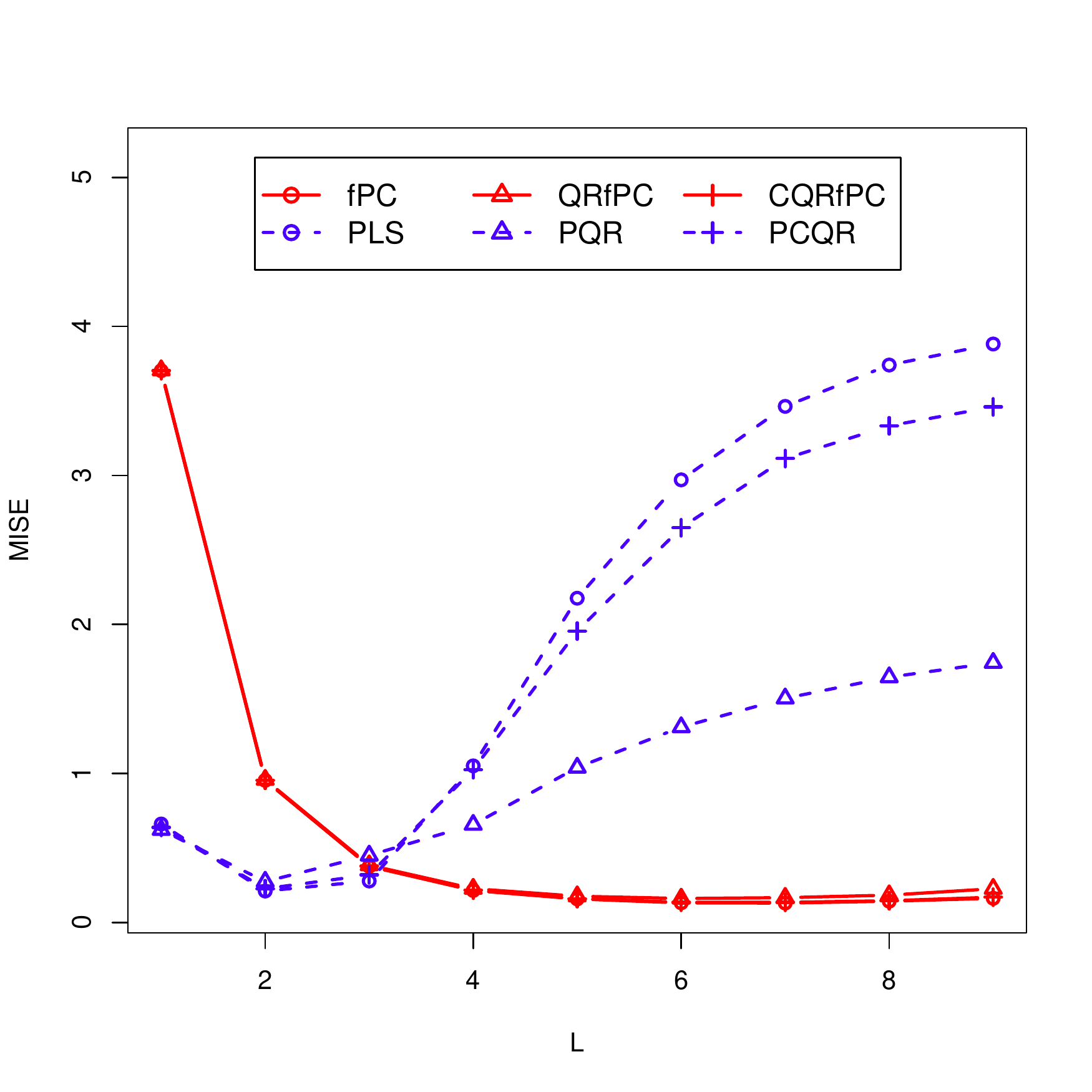}
 \end{minipage}
   \begin{minipage}[t]{.49\textwidth}
     \includegraphics[scale=0.385]{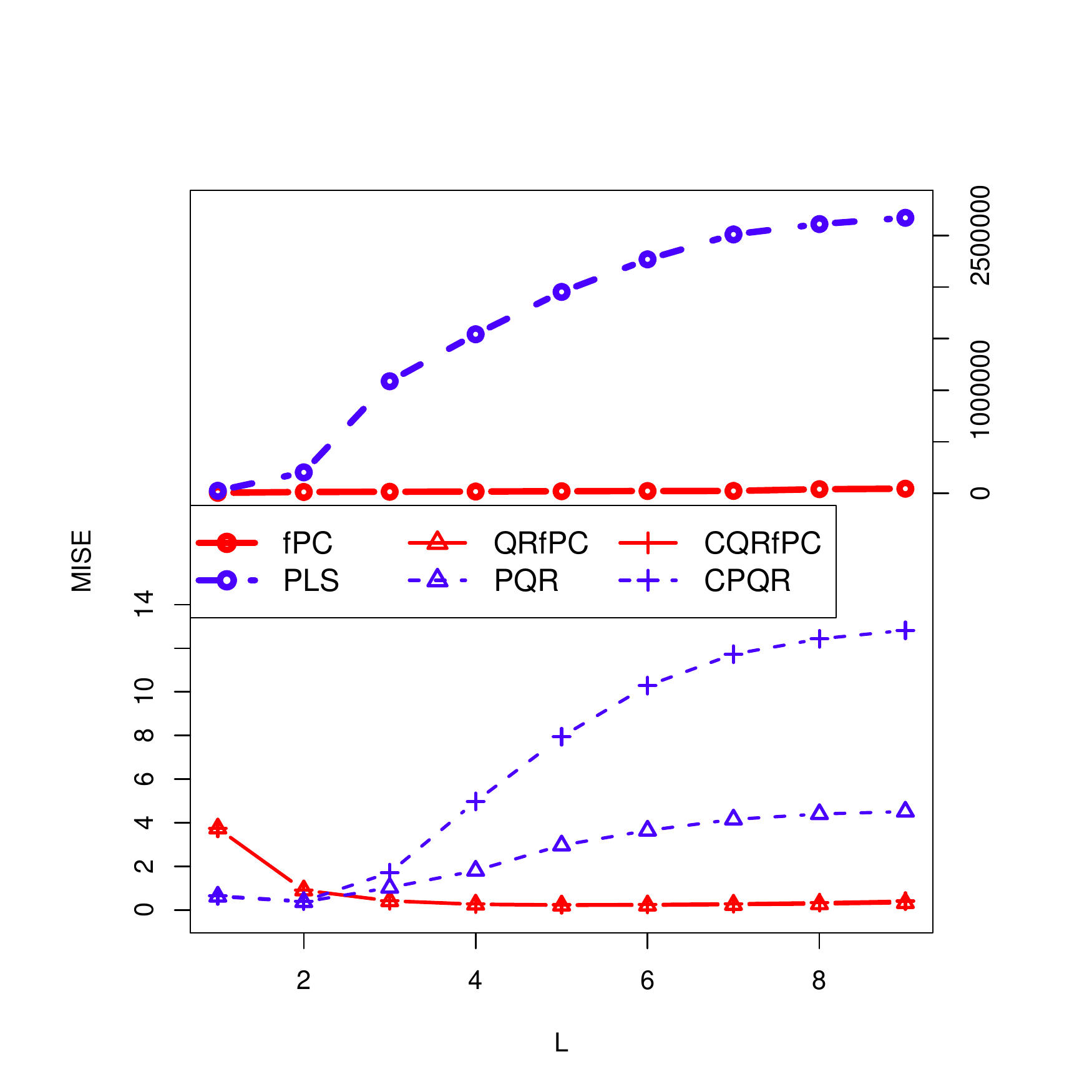}
 \end{minipage}
 \caption{Simulation I: the MISEs with Gaussian (left) and Cauchy (right) errors, sample size $n=100$, $200$, and $500$ from up to down.}
 \label{sim1f1}
\end{figure}

In the simulation, we set the total number of replication $S=100$. For the first three cutoff levels,
Table \ref{sim1t1} gives us a summary of the different configurations of parameters for the six methods.
Although the simulation design is in favour of fPC based methods,
for the small number of cutoff levels, the PLS, PQR and PCQR methods perform better regarding the performance measurements of $\mbox{Bias}^2$ and MISE. 
Due to the natures of sensitivity against skewness of errors, Figure
\ref{sim1f1} shows that the performances of PLS and fPC are much worse
in general compared with the other four methods when the errors follow
the Cauchy distribution.
On the other hand, when the Gaussian errors are employed, for the lower cutoff levels, the PLS, PQR and PCQR  methods are very similar.
And when the number of cutoff levels becomes larger, the PCQR performs slightly better than the PLS while PQR performs much better than the PCQR.
fPC based methods are similar to each other crossing all cutoff levels.

\begin{figure}[htbp]
 \centering
  \begin{minipage}[t]{.49\textwidth}
     \includegraphics[scale=0.36]{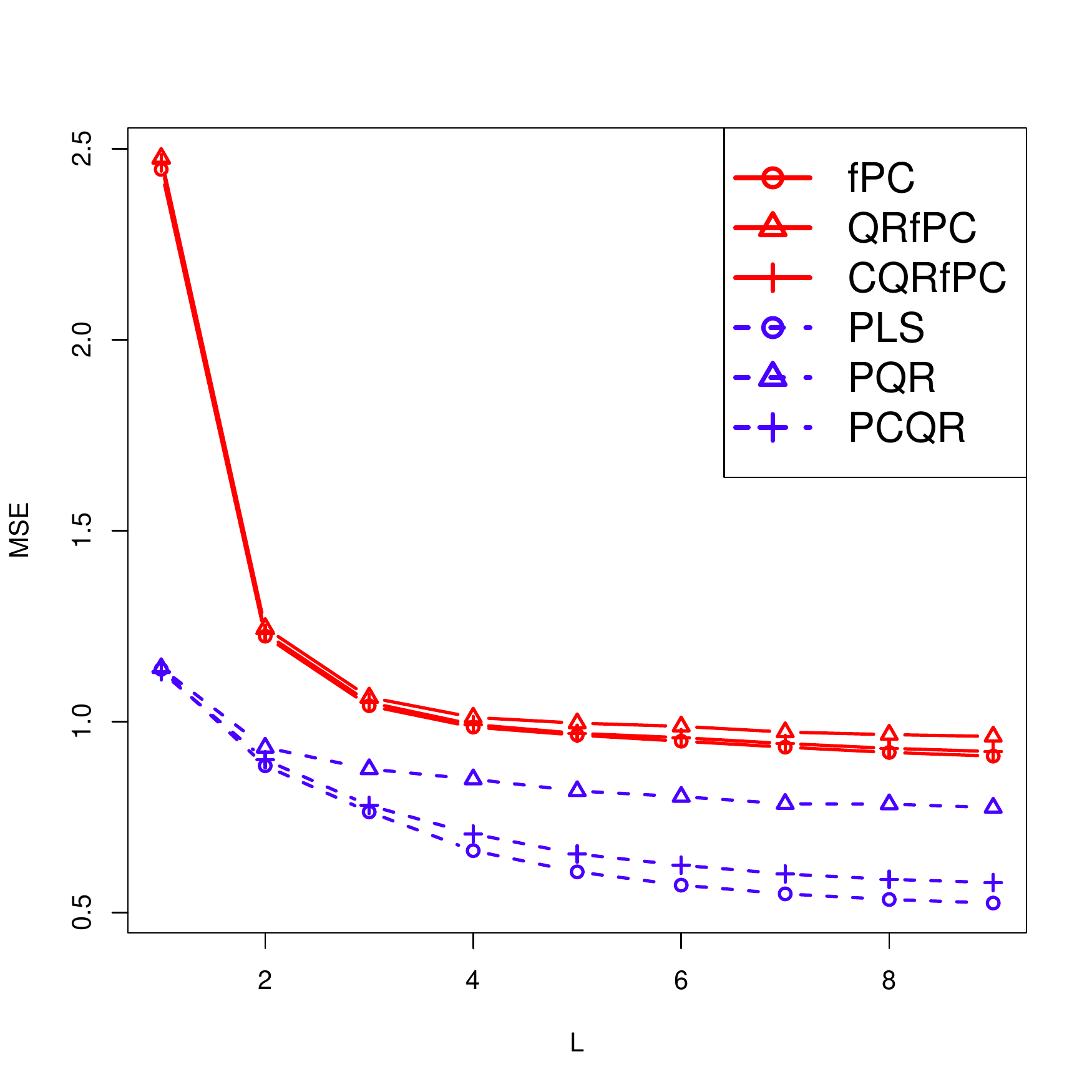}
 \end{minipage}
  \begin{minipage}[t]{.49\textwidth}
     \includegraphics[scale=0.385]{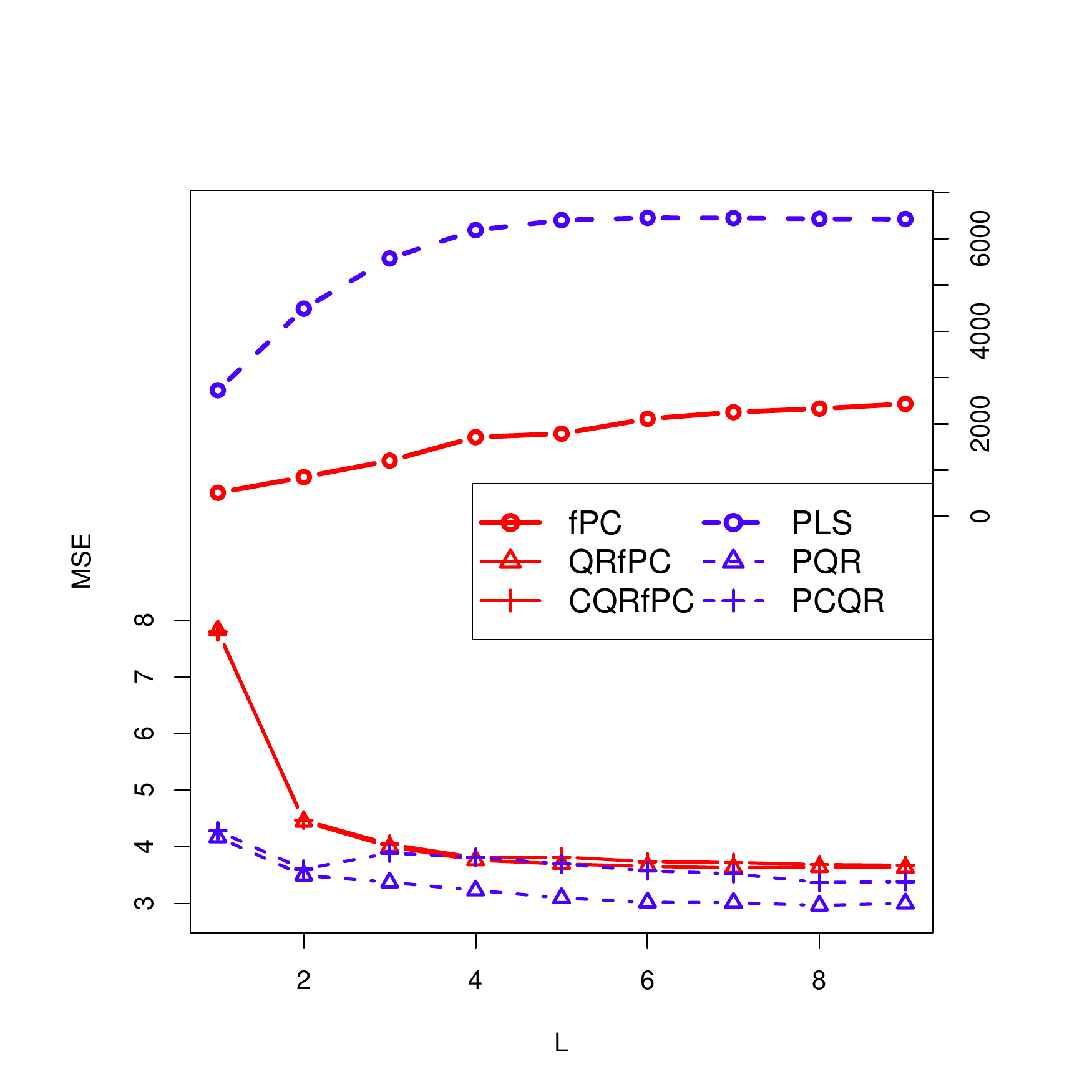}
 \end{minipage}
   \begin{minipage}[t]{.49\textwidth}
     \includegraphics[scale=0.36]{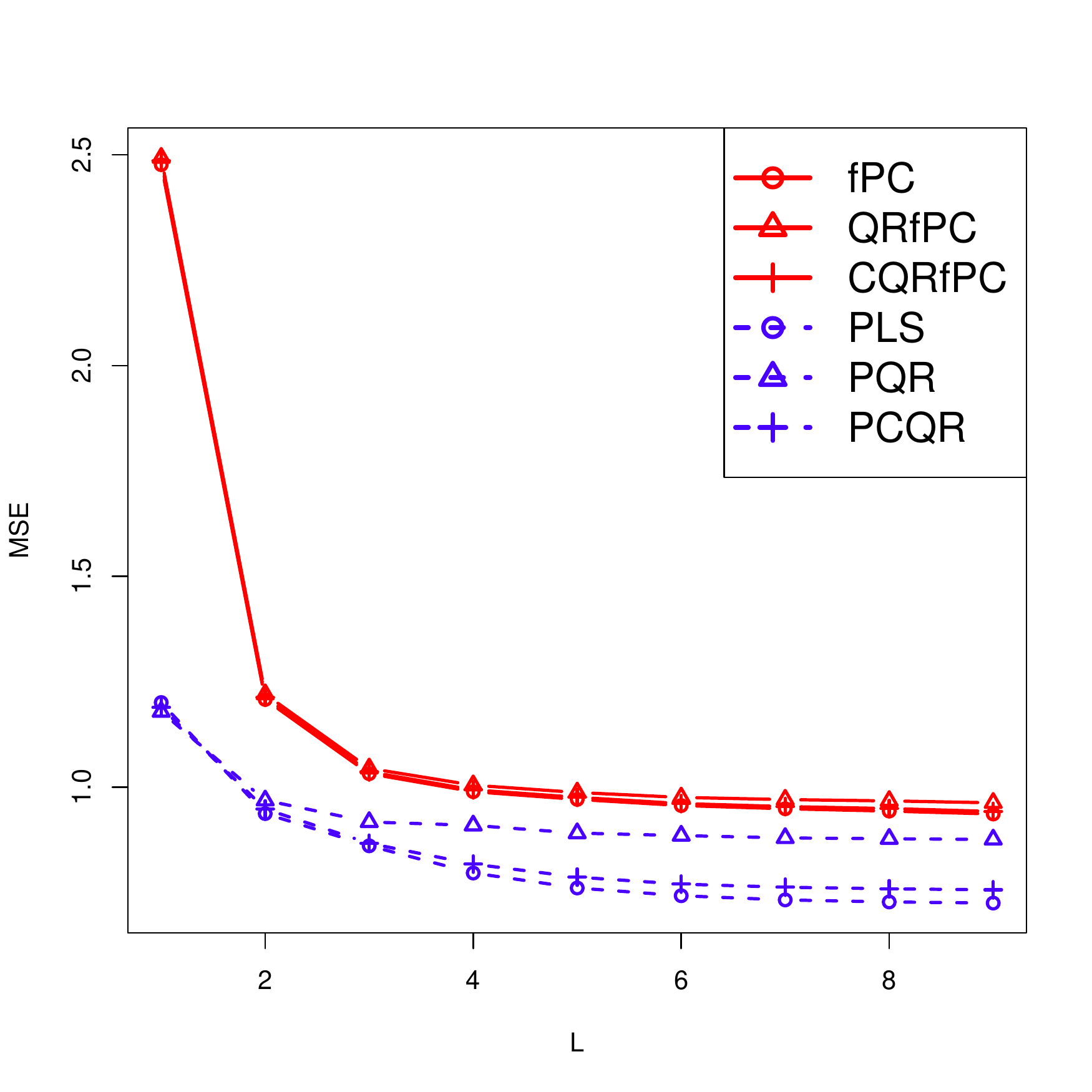}
 \end{minipage}
   \begin{minipage}[t]{.49\textwidth}
     \includegraphics[scale=0.385]{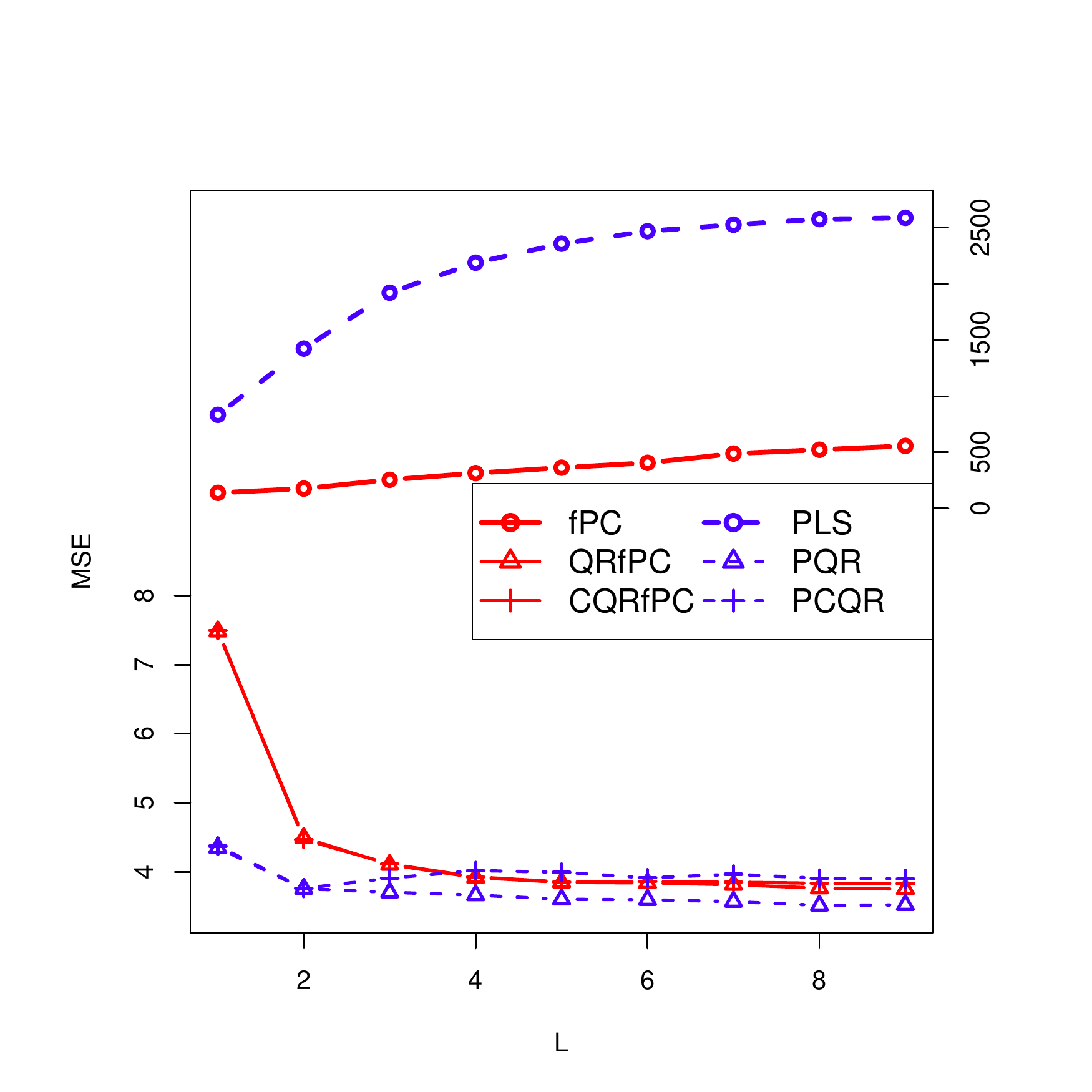}
 \end{minipage}
  \begin{minipage}[t]{.49\textwidth}
     \includegraphics[scale=0.36]{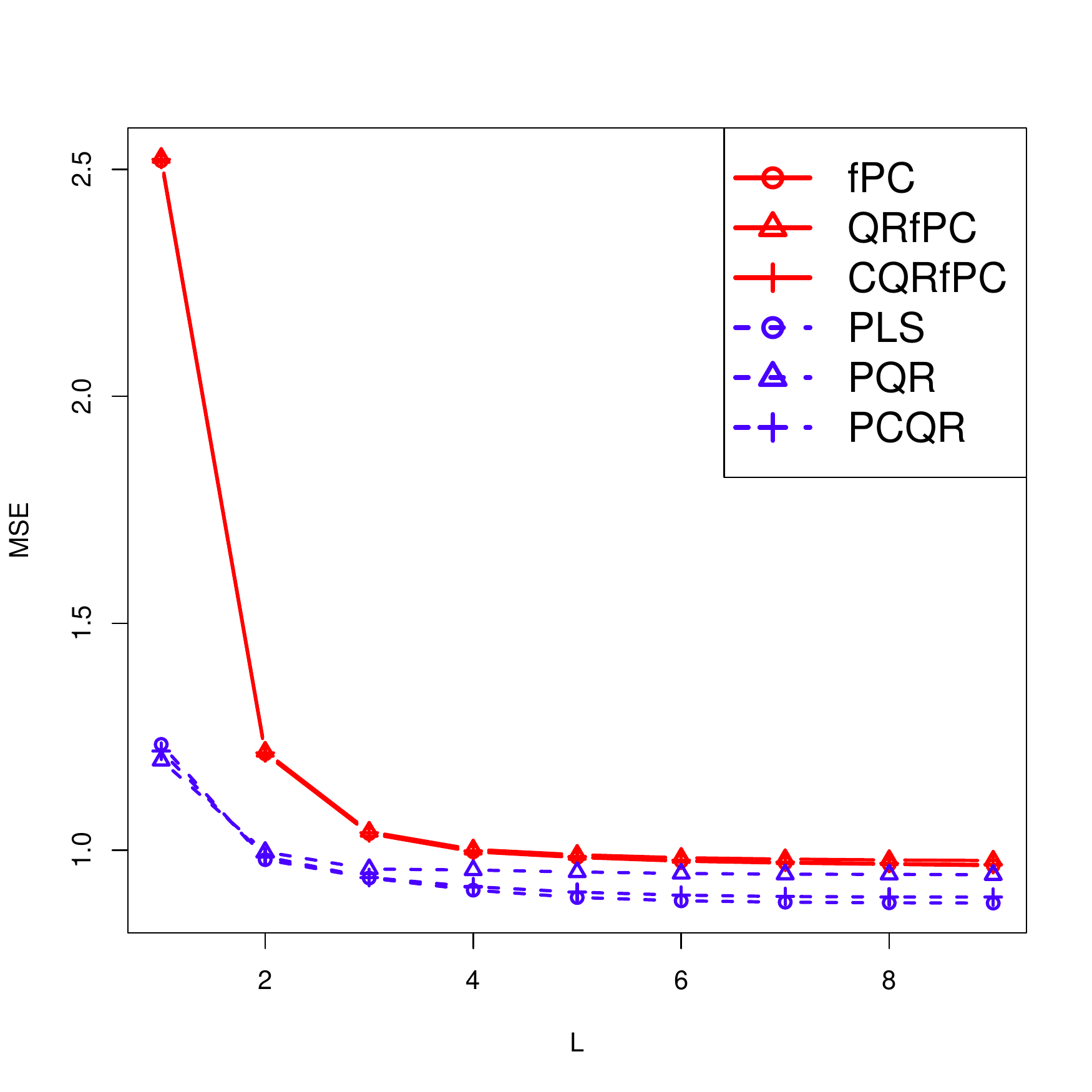}
 \end{minipage}
   \begin{minipage}[t]{.49\textwidth}
     \includegraphics[scale=0.385]{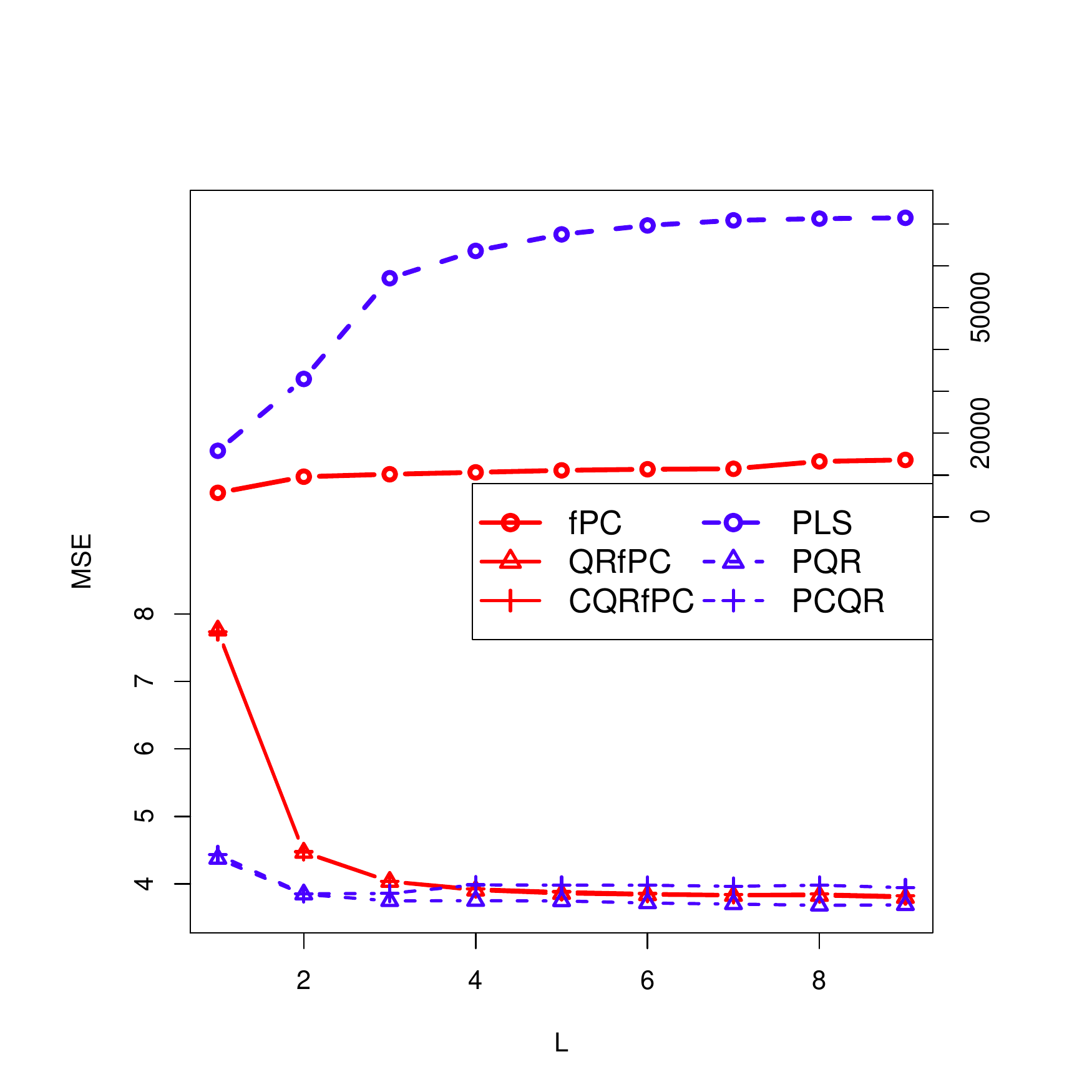}
 \end{minipage}
 \caption{Simulation I: the averaged MSEs with Gaussian (left) and Cauchy (right) errors, sample size $n=100$, $200$, and $500$ from up to down.}
 \label{sim1f2}
\end{figure}

The averaged mean squared error (MSE) of the responses is another
prediction performance criteria we consider. Figure \ref{sim1f2}
indicates that the prediction errors are much lower for PLS, PQR, PCQR
methods compared with those for fPC based methods, due to the fact
that fPC based methods are only data driven while the other three
methods are both data and response adapted. For the Gaussian errors,
although with regard to the functional coefficients estimation PQR is
better than both PLS and PCQR methods, taking into consider of the
prediction errors, the PLS and PCQR methods perform better than PQR.
For the Cauchy errors, PQR performs the best out of the PLS, PQR, PCQR
methods which indicates that PQR is more robust against the skewness
of error distribution.

{\sl Simulation II.}  In this simulation, we take the $Z_i(t)$s from a real data study, and generate the $Y_i$s according to the linear model of 
$$Y = \int_{0}^{1} \gamma(t) Z(t) dt + \varepsilon,$$
where the error $\varepsilon$ is taken as Gaussian and Cauchy.
The centres of errors are taken as zero while the scales are taken as the empirical standard deviation of the true responses multiplied by $\sqrt{5}$.
The $Z_i$s are taken from a benchmark Phoneme dataset, which can be downloaded from {\sl http://statweb.stanford.edu\\/~tibs/ElemStatLearn/}.
In these data, $Z_i(t)$ represents log-periodgrams constructed from recordings of different phonemes.
The periodgrams are available at $256$ equally-spaced frequencies $t$,
which for simplicity we denote by $0=t_1<t_1<\ldots<t_{m}=1$, where $m=256$ \cite{hastie2009elements}.
We used $n=1717$ data curves $Z_i(t)$ that correspond to the phonemes ``aa'' as in ``dark'' and ``ao'' as in ``water''.
This example can also be found in \cite{delaigle2012methodology}.

Computing the first $J=20$ empirical fPCbasis functions $\hat{\phi}_1(t),\ldots,\hat{\phi}_{J}(t)$, we consider four different curves $\gamma(t)$ by taking
$\gamma(t)=\sum_{j=1}^J a_j \hat{\phi}_i(t)$ for four different sequence of $a_j$s: (i) $a_j=(-1)^j \cdot \mathbf{1}\{0\leq j\leq 5\}$;
(ii) $a_j=(-1)^j \cdot \mathbf{1}\{6\leq j\leq 10\}$; (iii) $a_j=(-1)^j \cdot \mathbf{1}\{11\leq j\leq 15\}$; (iv) $a_j=(-1)^j \cdot \mathbf{1}\{16\leq j\leq 20\}$.
Going through case (i) to (iv), the models become less favorable for fPC, while we will see the PLS, PQR and PCQR methods manage to capture the interaction
between $Z$ and $Y$ using only a few terms.

\begin{figure}[htbp]
 \centering
  \begin{minipage}[t]{.35\textwidth}
     \includegraphics[scale=0.26]{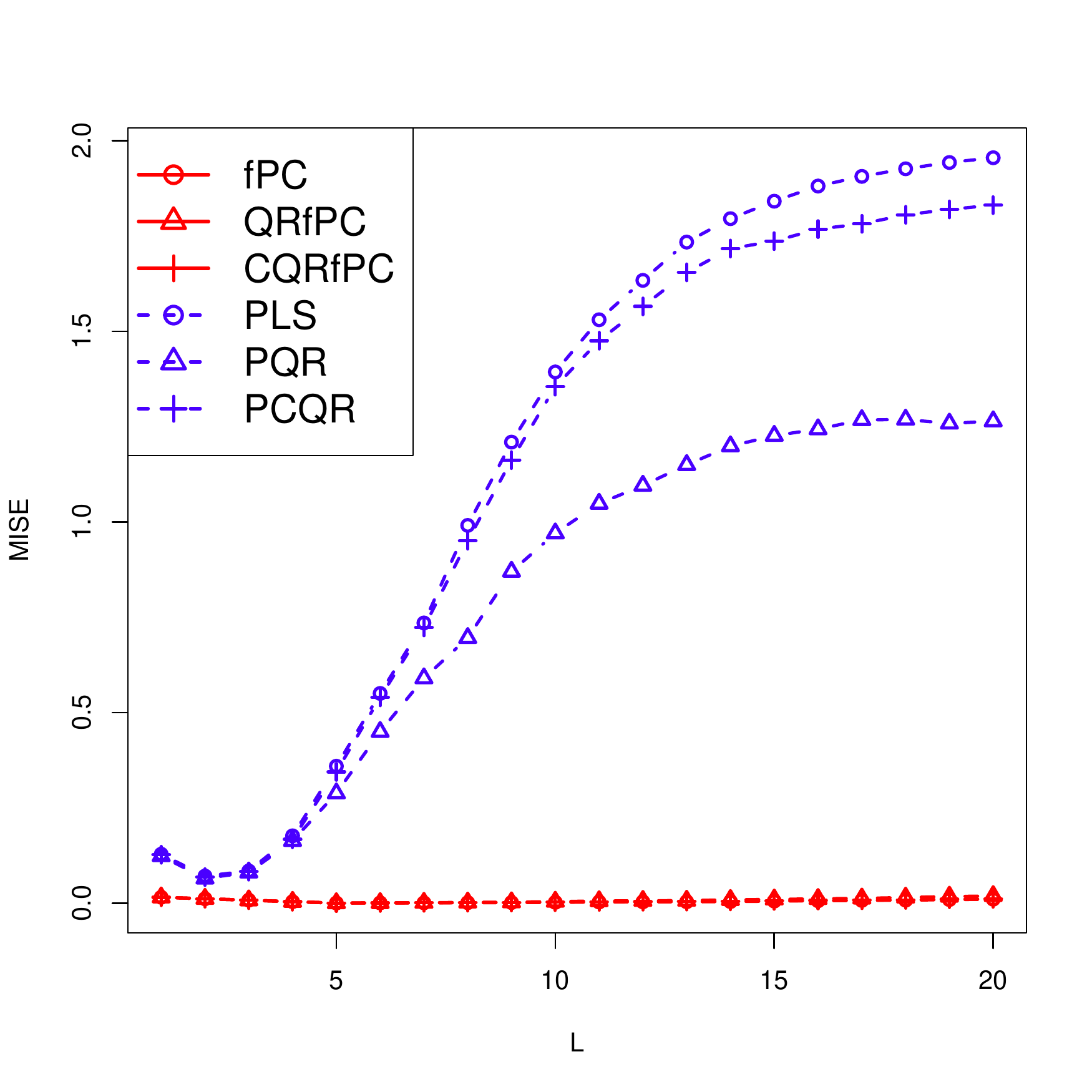}
 \end{minipage}
  \begin{minipage}[t]{.35\textwidth}
     \includegraphics[scale=0.28]{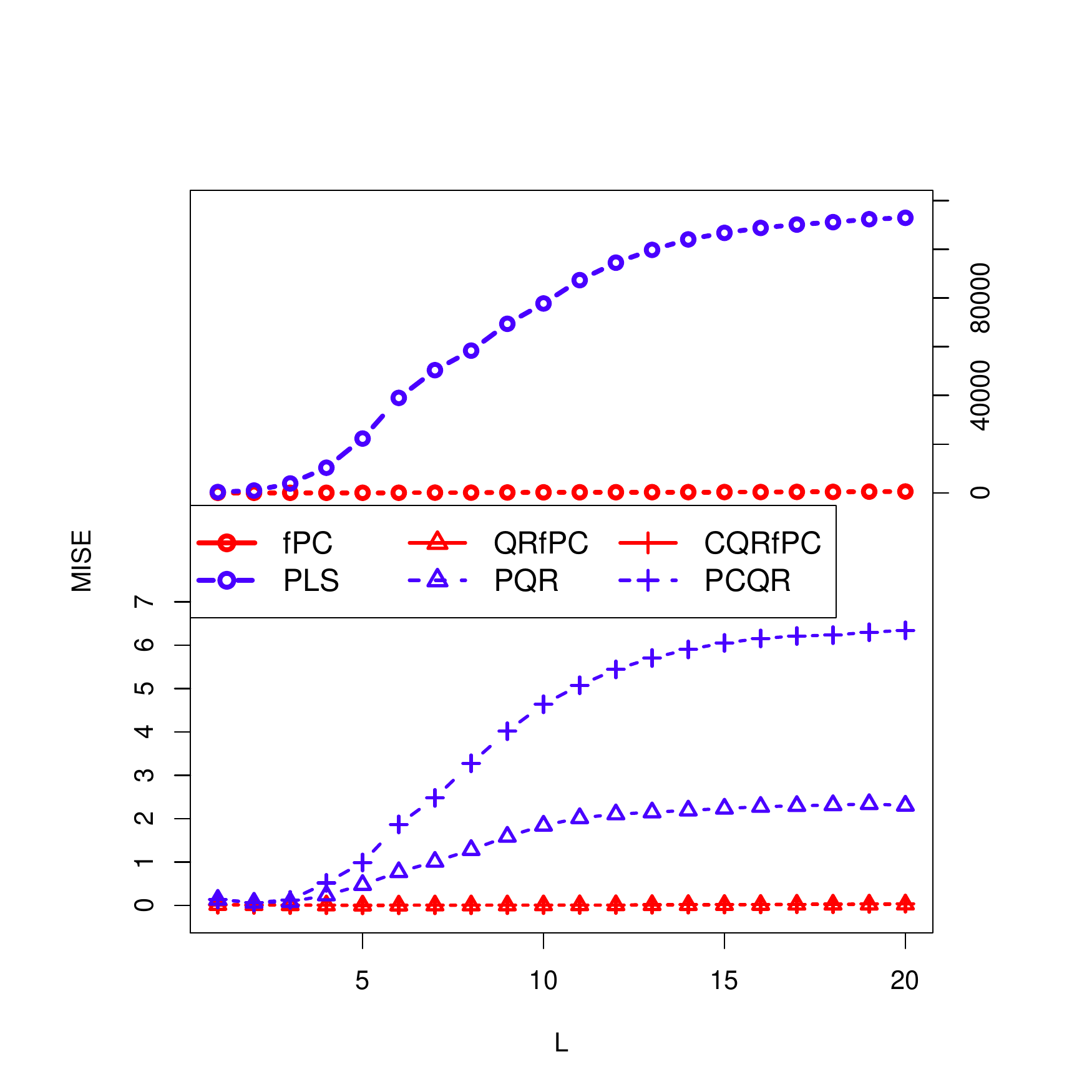}
 \end{minipage}
   \begin{minipage}[t]{.35\textwidth}
     \includegraphics[scale=0.26]{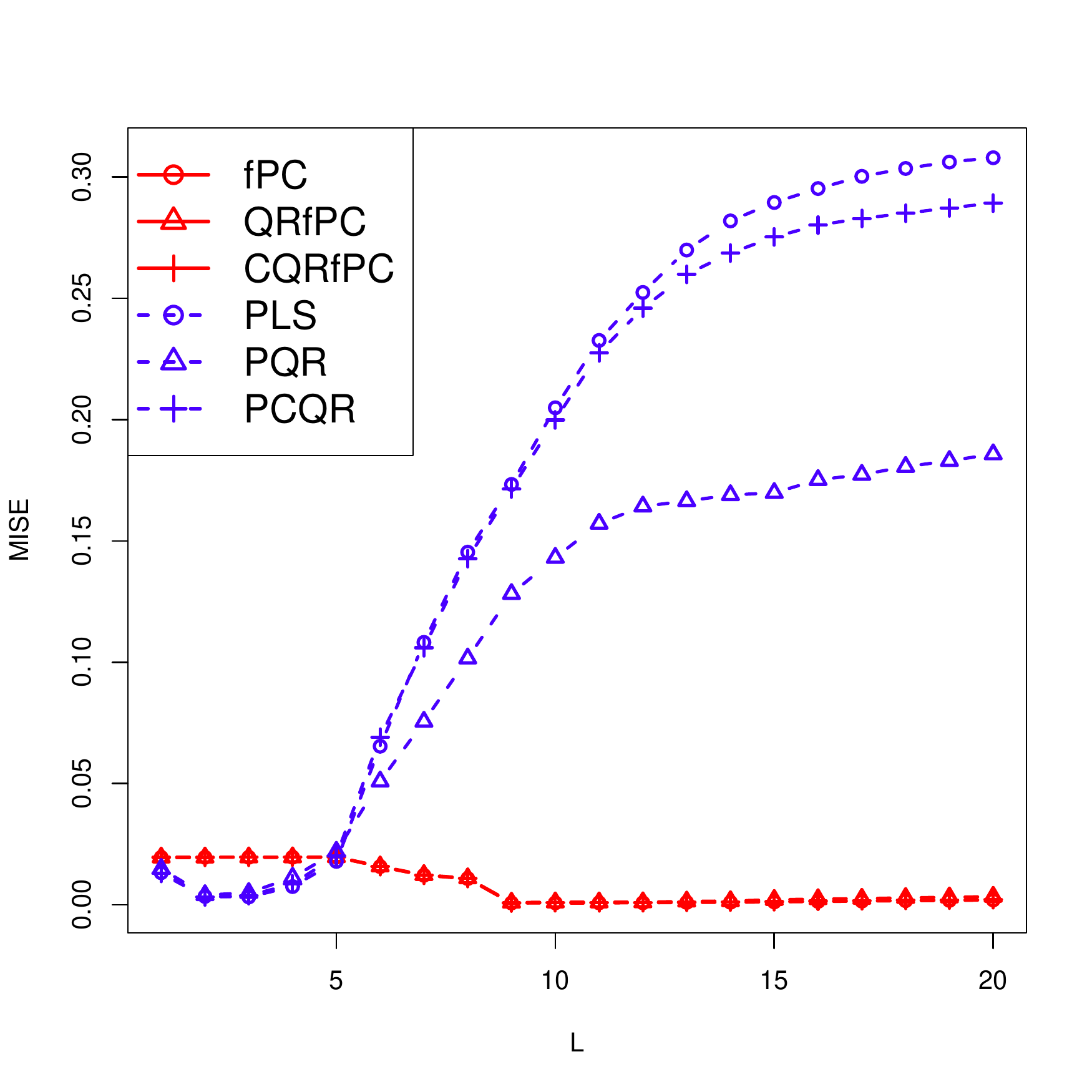}
 \end{minipage}
   \begin{minipage}[t]{.35\textwidth}
     \includegraphics[scale=0.28]{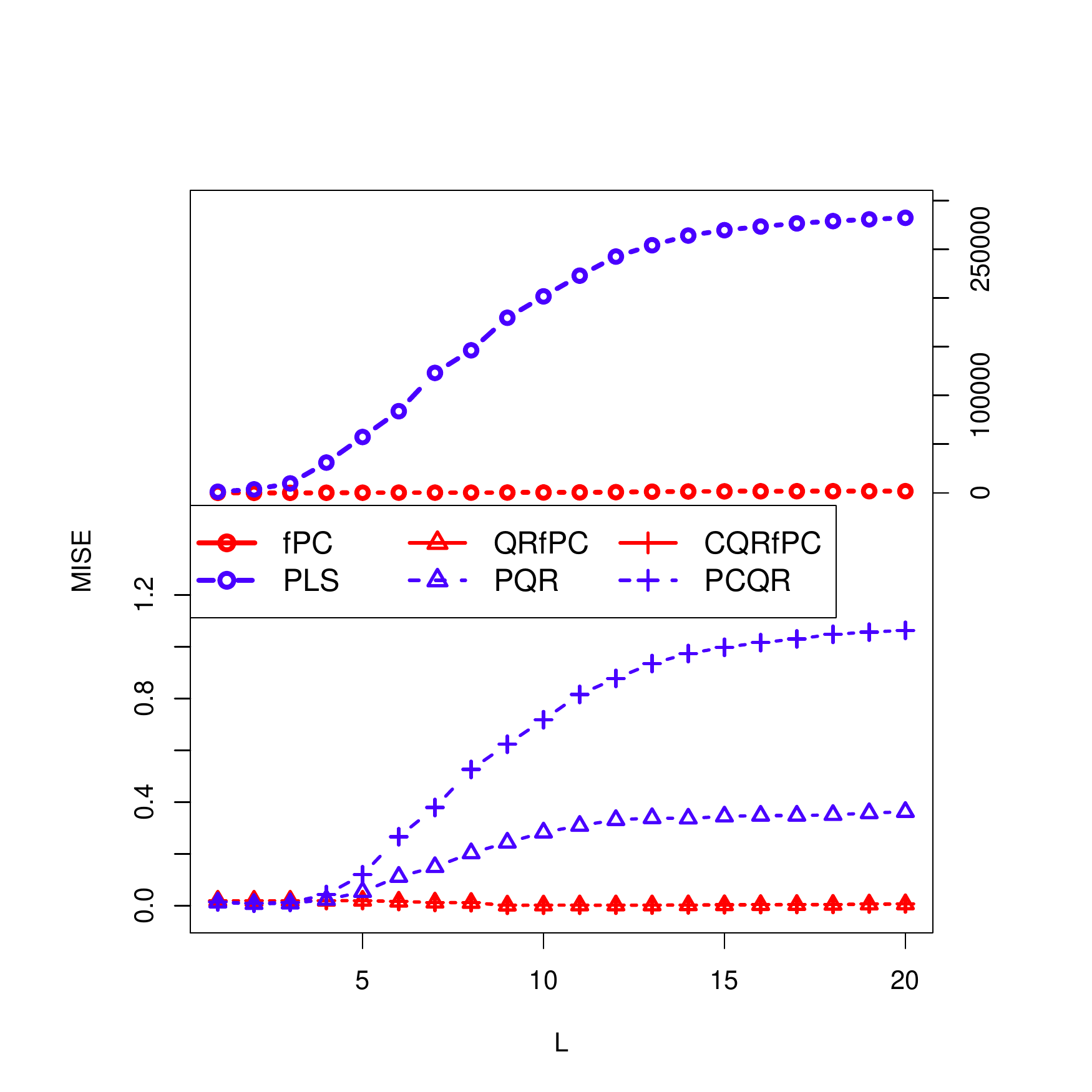}
 \end{minipage}
   \begin{minipage}[t]{.35\textwidth}
     \includegraphics[scale=0.26]{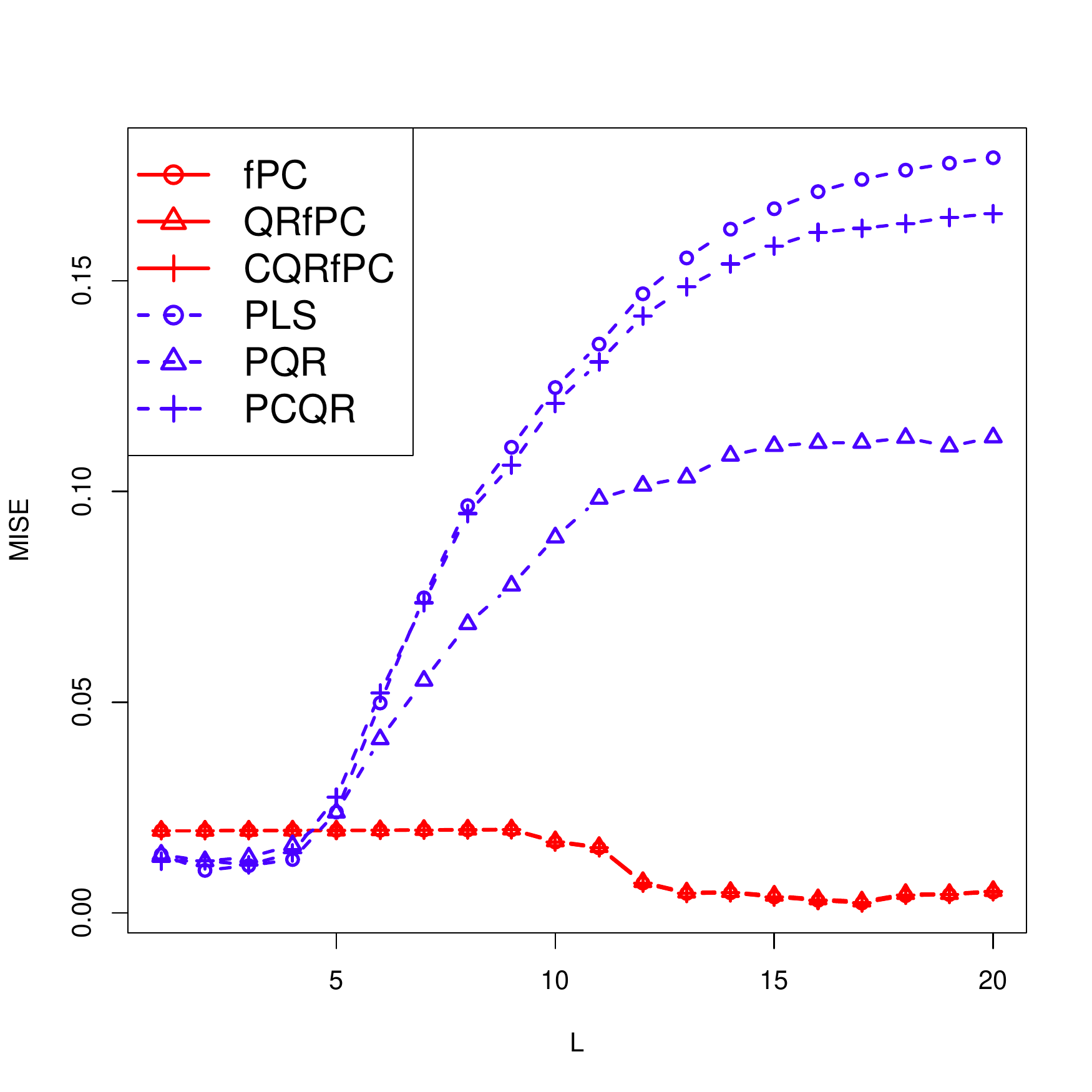}
 \end{minipage}
  \begin{minipage}[t]{.35\textwidth}
     \includegraphics[scale=0.28]{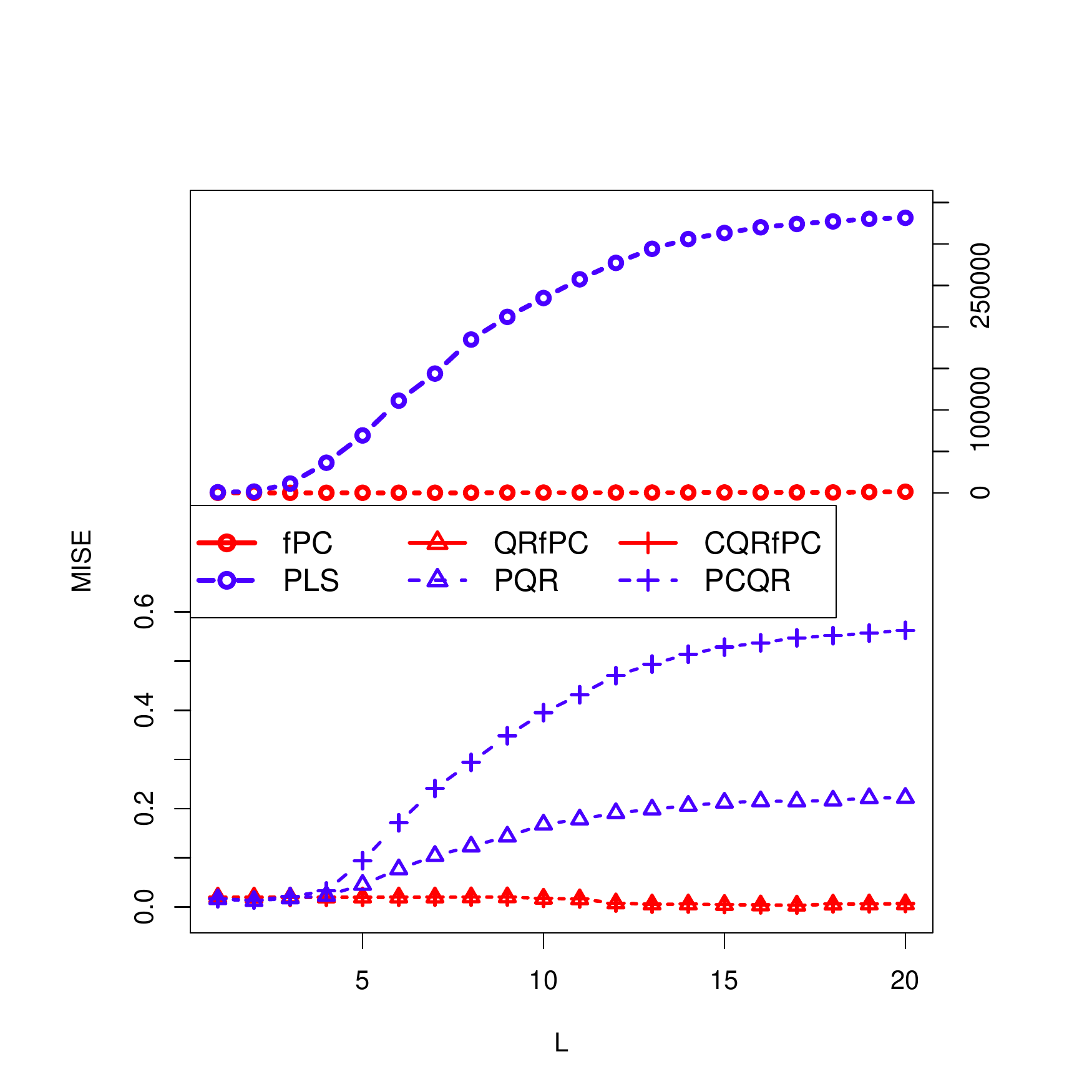}
 \end{minipage}
   \begin{minipage}[t]{.35\textwidth}
     \includegraphics[scale=0.26]{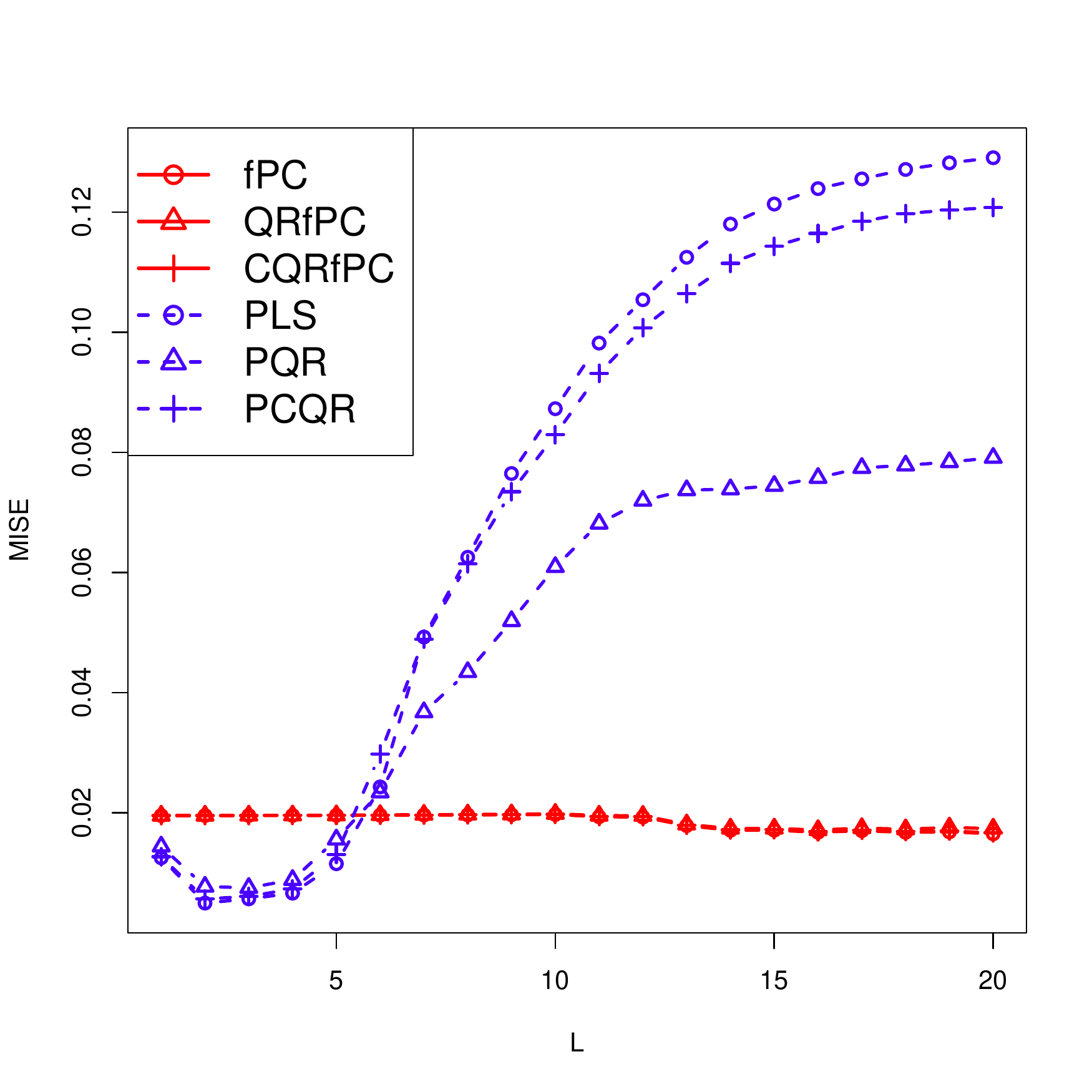}
 \end{minipage}
   \begin{minipage}[t]{.35\textwidth}
     \includegraphics[scale=0.28]{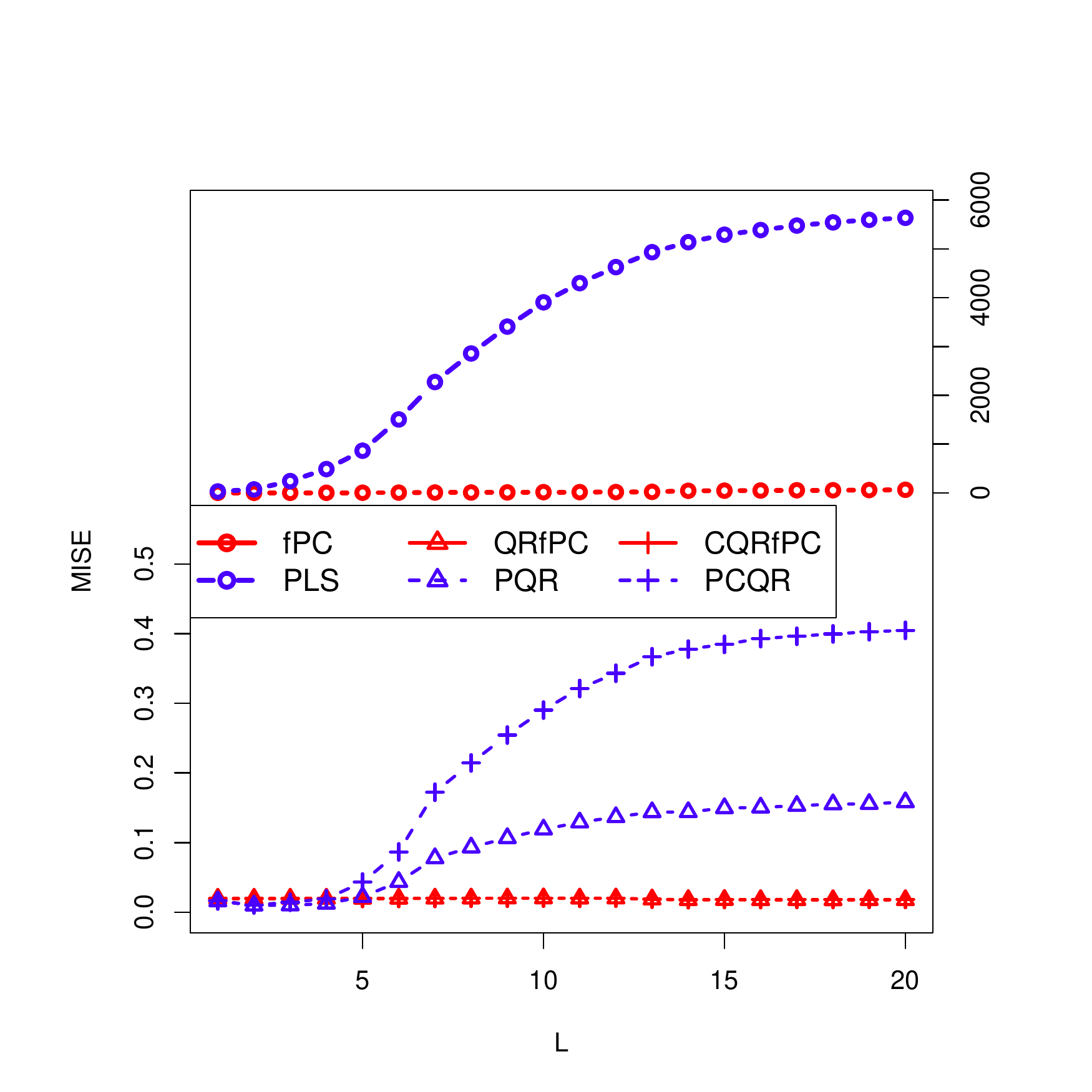}
 \end{minipage}
 \caption{Simulation II: the MISEs with Gaussian (left) and Cauchy (right) errors, case I, II, II, and IV from up to down.}
 \label{sim2f4}
\end{figure}

We take $\tau = 0.5$ and compare the six methods by looking at the $\mbox{MISE}$, $\mbox{Bias}^2$ and $\mbox{Var}$. As shown in Figure \ref{sim2f4}, 
from case (i) to (iv), PLS, PQR and PCQR methods perform better and better compared with the fPC based methods.
In fact, all the PLS, PQR and PCQR  methods manage to obtain a very good fitting using only a much small number of components no matter how the errors are distributed. 
This shows great superiority of our proposed methods when the functional coefficients do not lie on the fPC space.

\begin{figure}[htbp]
 \centering
  \begin{minipage}[t]{.35\textwidth}
     \includegraphics[scale=0.26]{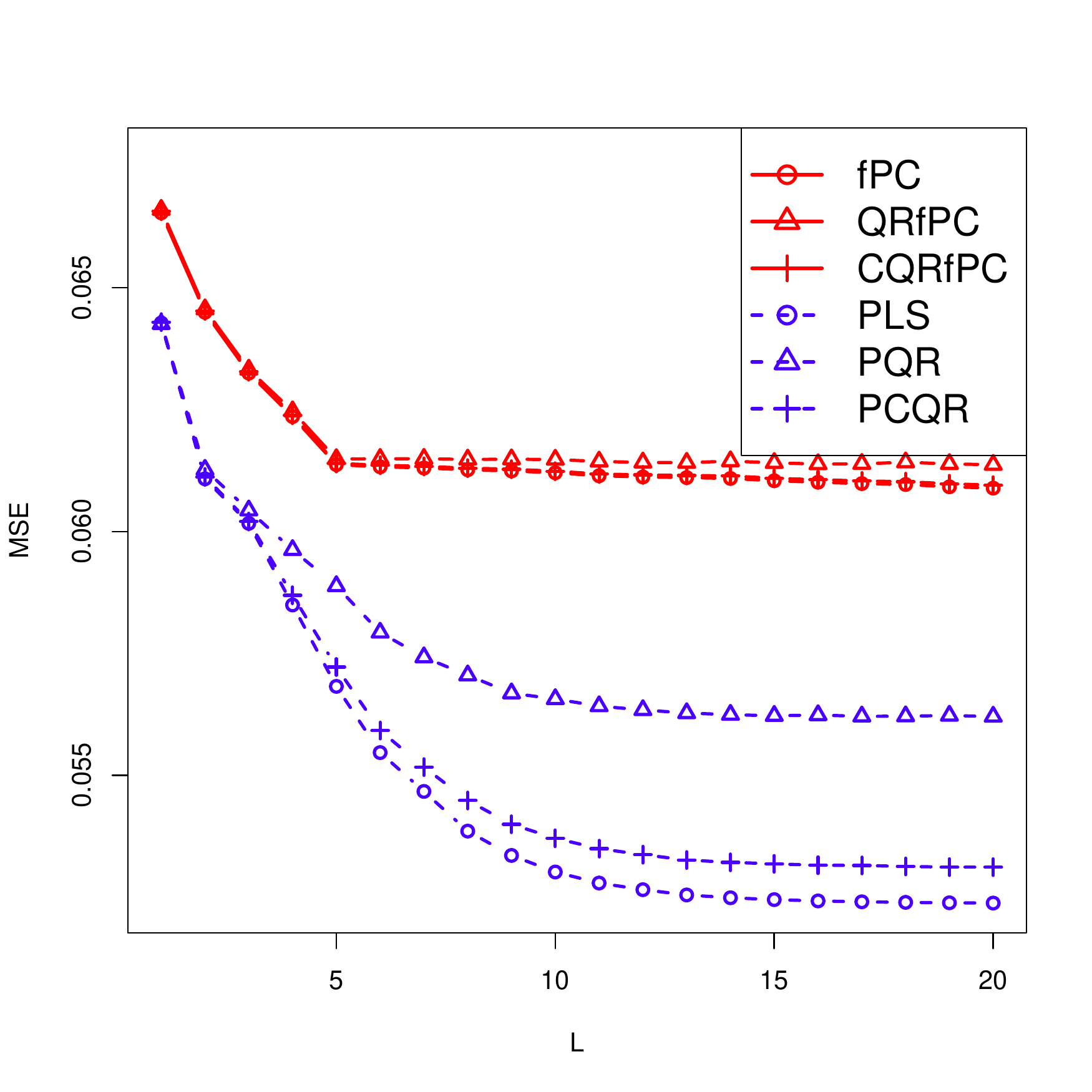}
 \end{minipage}
  \begin{minipage}[t]{.35\textwidth}
     \includegraphics[scale=0.28]{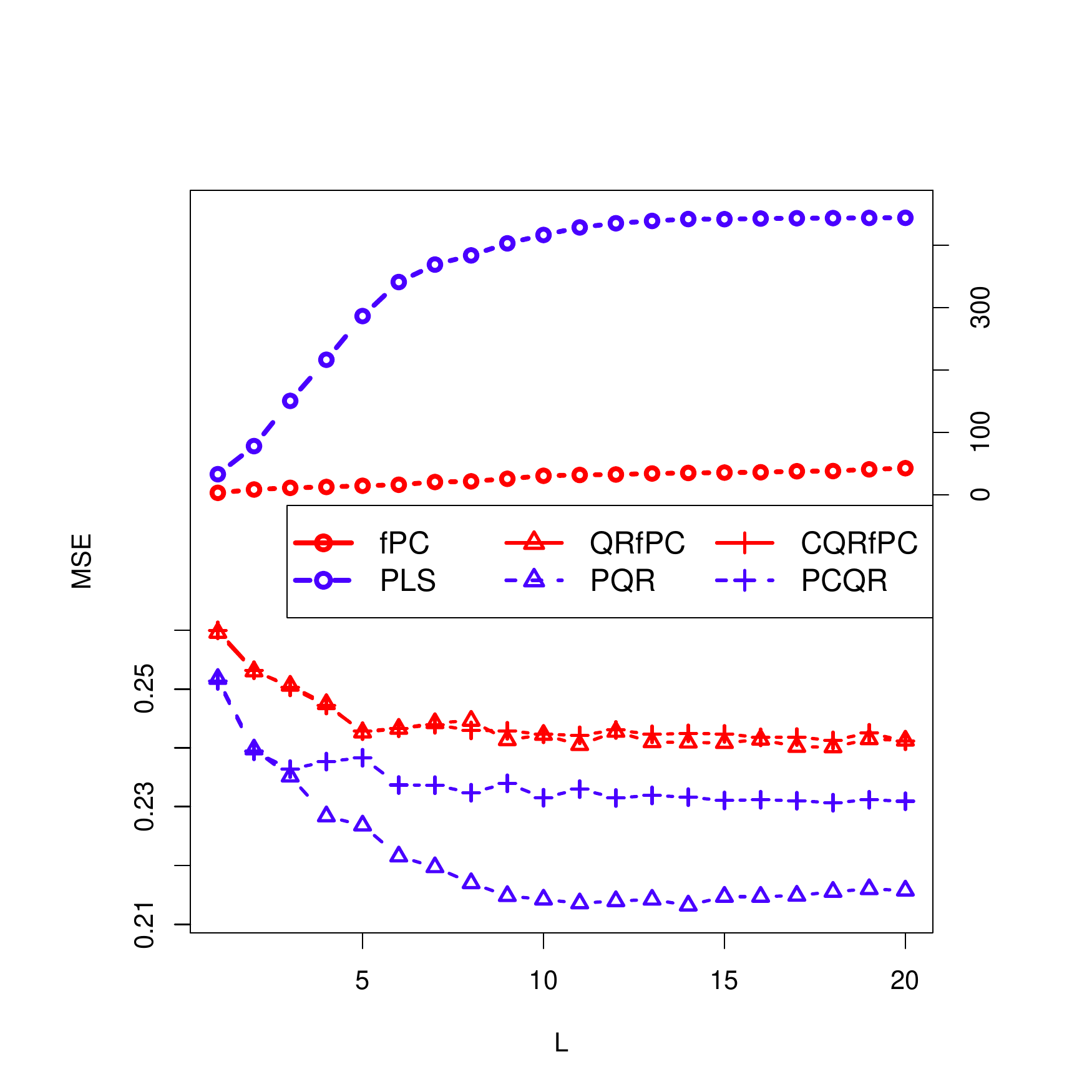}
 \end{minipage}
   \begin{minipage}[t]{.35\textwidth}
     \includegraphics[scale=0.26]{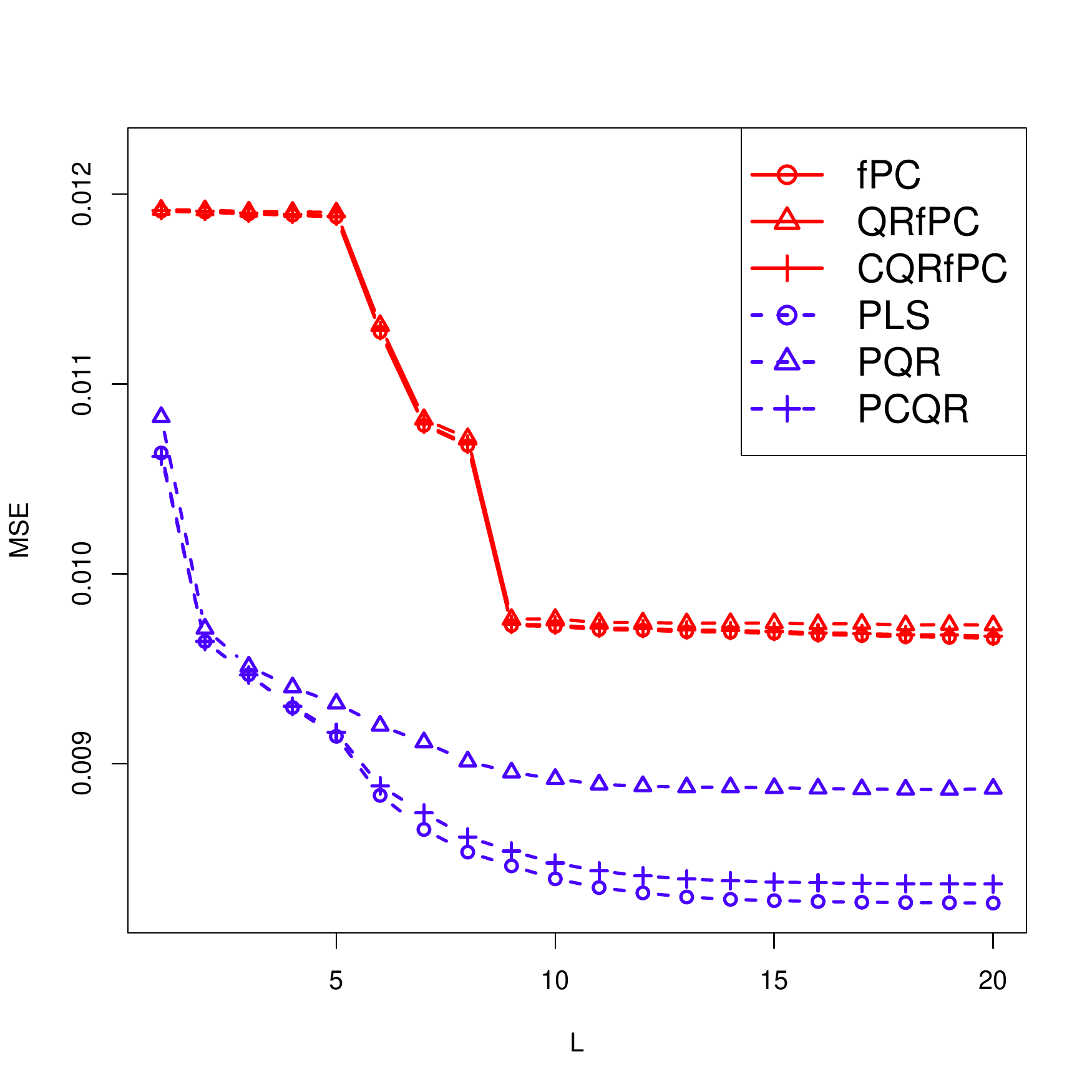}
 \end{minipage}
   \begin{minipage}[t]{.35\textwidth}
     \includegraphics[scale=0.28]{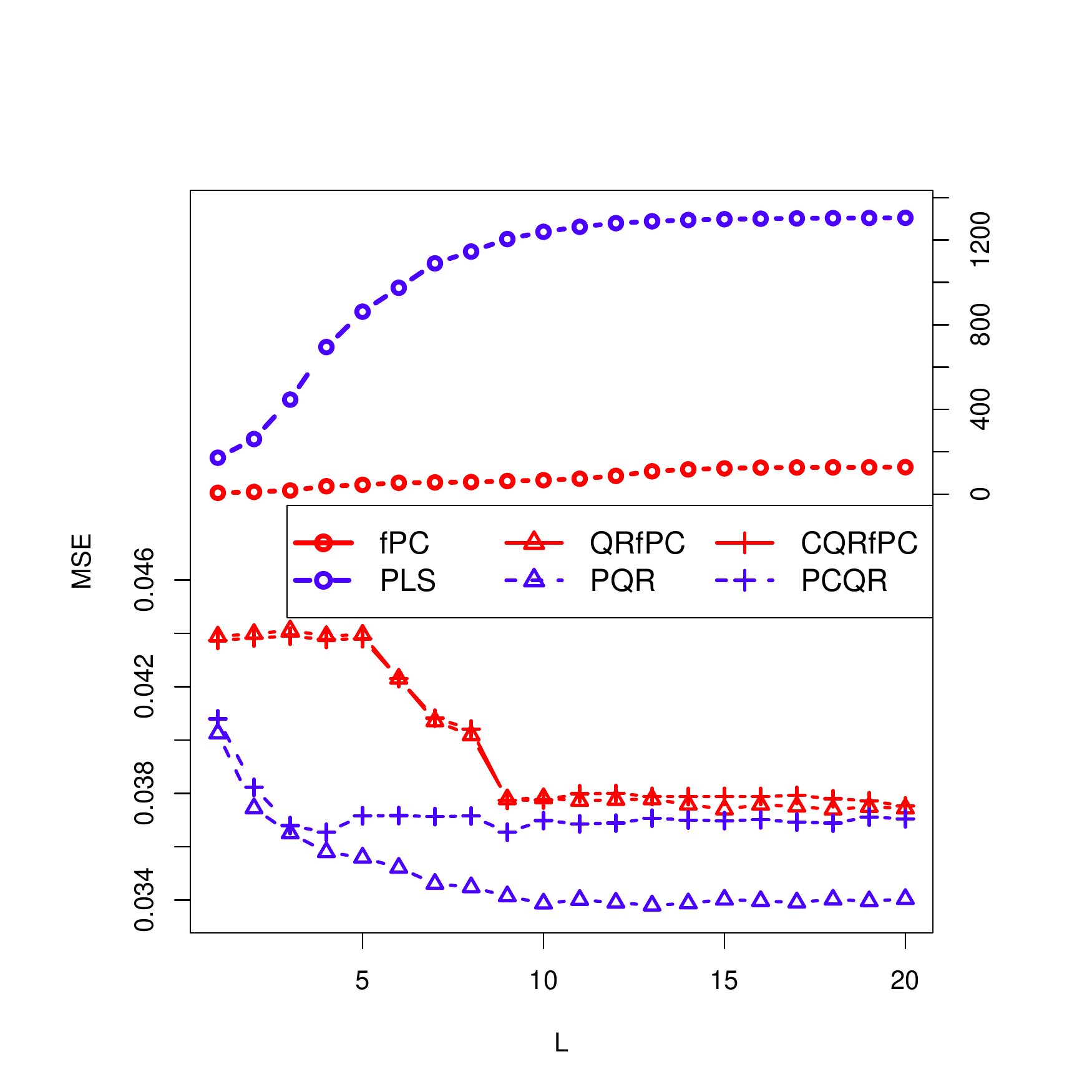}
 \end{minipage}
   \begin{minipage}[t]{.35\textwidth}
     \includegraphics[scale=0.26]{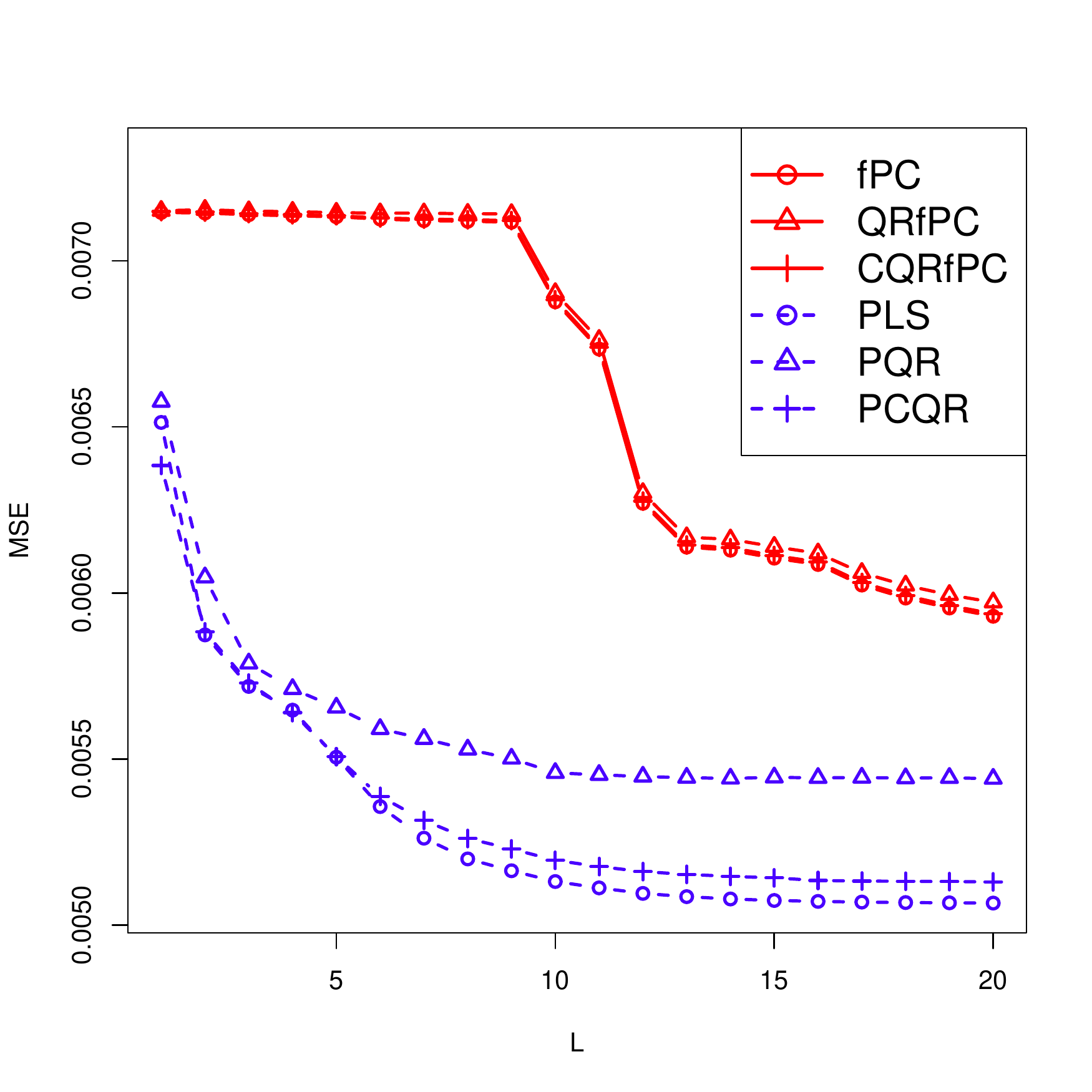}
 \end{minipage}
  \begin{minipage}[t]{.35\textwidth}
     \includegraphics[scale=0.28]{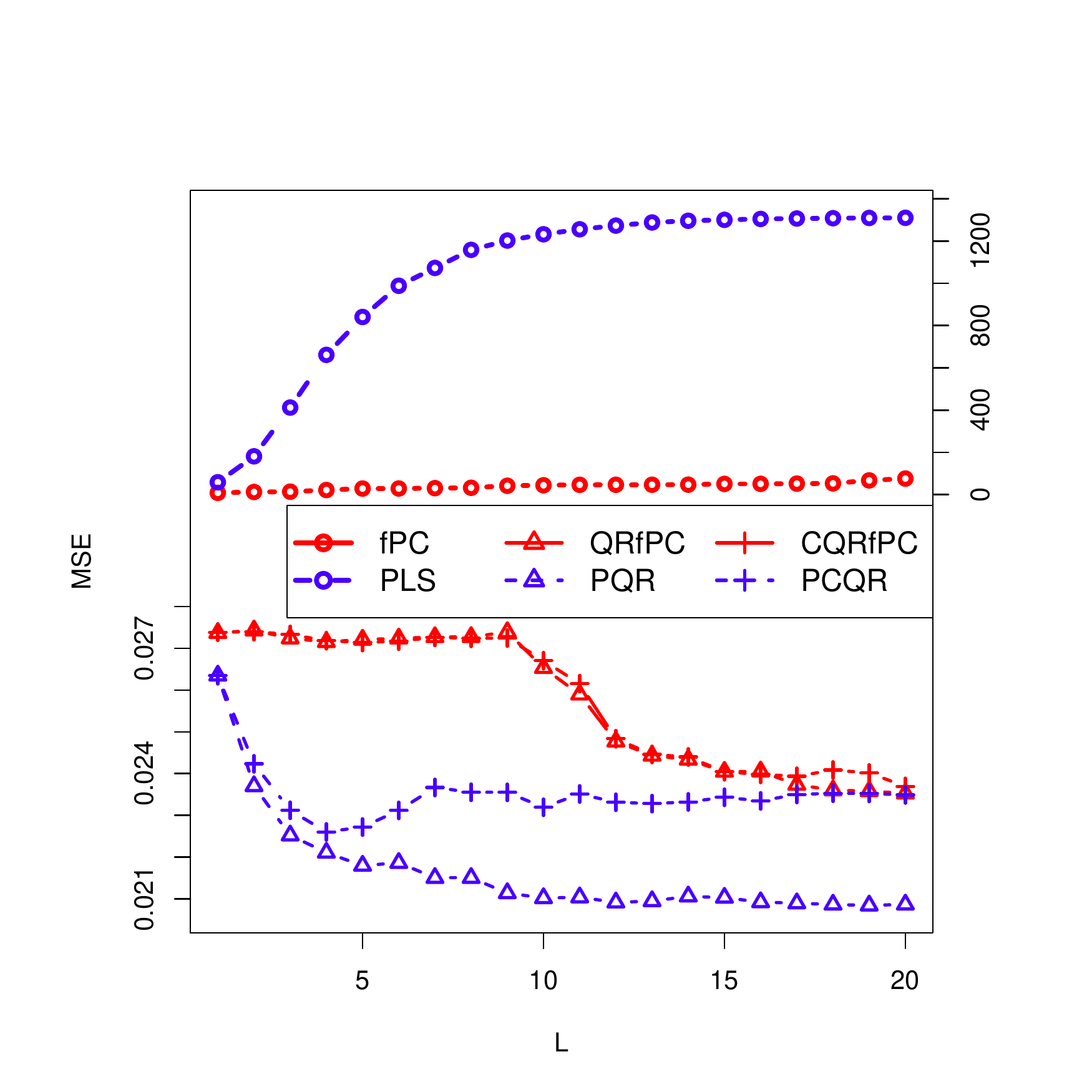}
 \end{minipage}
   \begin{minipage}[t]{.35\textwidth}
     \includegraphics[scale=0.26]{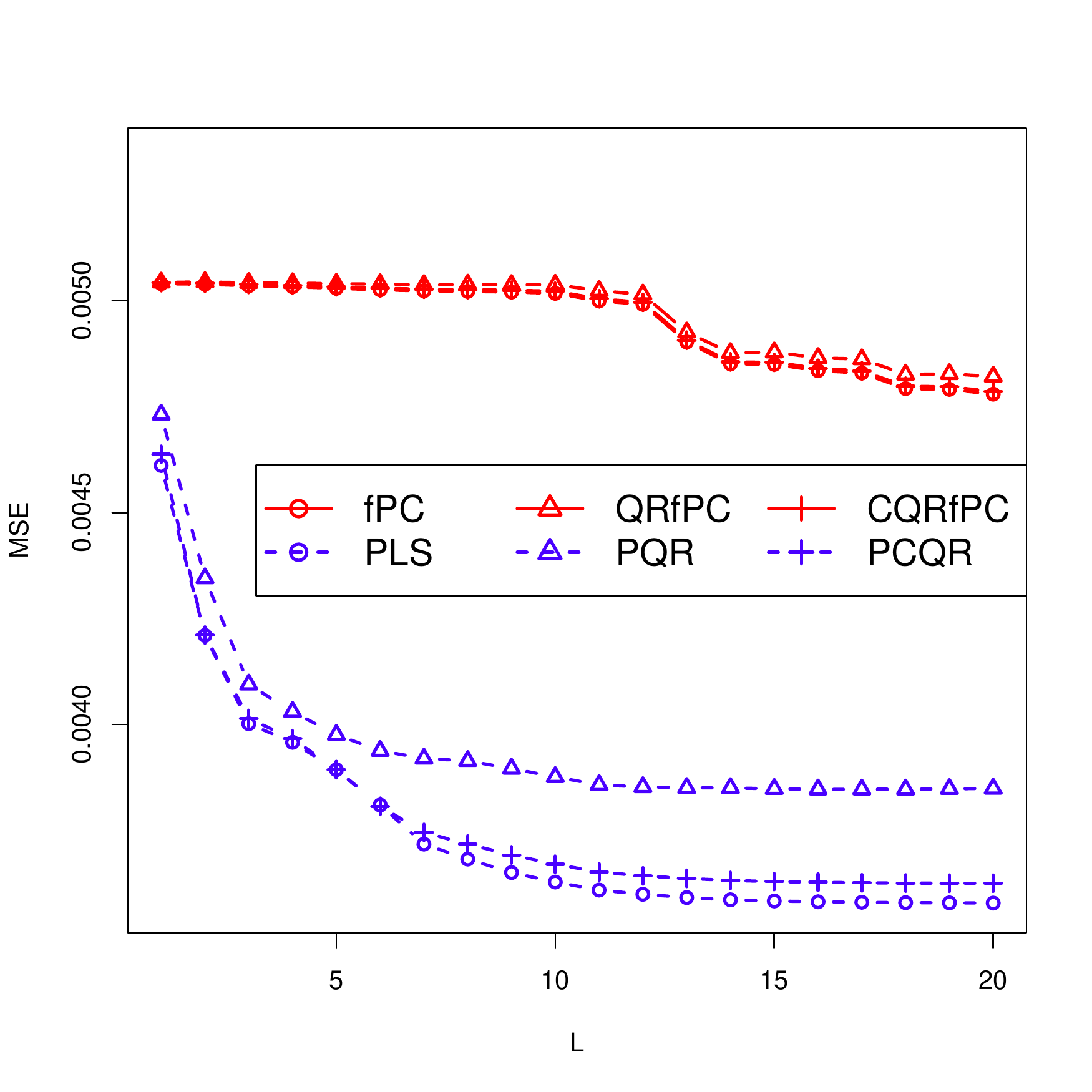}
 \end{minipage}
   \begin{minipage}[t]{.35\textwidth}
     \includegraphics[scale=0.28]{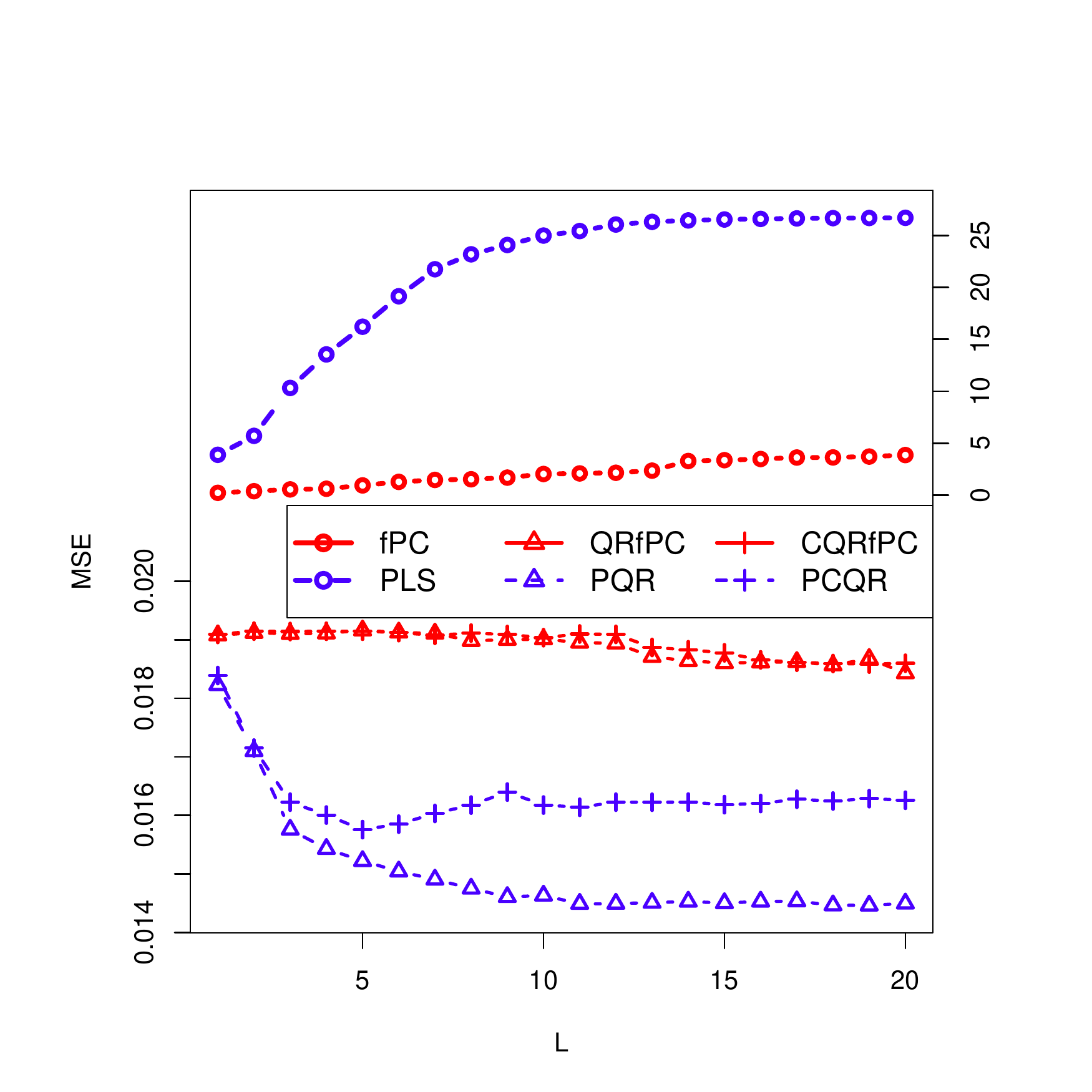}
 \end{minipage}
 \caption{Simulation II: the averaged MSEs with Gaussian (left) and Cauchy (right) errors, case I, II, II, and IV from up to down.}
 \label{sim2f5}
\end{figure}

Figure \ref{sim2f5} displays the prediction errors MSE when the errors follow Gaussian (left panels) and Cauchy (right panels) distributions.
The PLS, PQR and PCQR  methods predict better in general compared with fPC based methods.
Except for the PLS of Cauchy errors, the MSEs of PQR and PCQR  methods decrease immediately with the increase of cutoff levels,
while the fPC based methods performed differently under each case.
From case (i) to (iv), the MSEs of fPC based methods begin to drop significantly after a larger and larger cutoff level.
And for the same cutoff levels, the differences of the prediction errors between the  PLS, PQR and PCQR  methods and the fPC based methods become more and more signifiant from case (i) to (iv).

One interesting phenomenon here is that although the PCQR method outperforms the PQR method when the errors are Gaussian distributed,
the PQR method regains its superiority when the errors are Cauchy distributed.
Compared with what we have observed from simulation I,
it may indicate that the PCQR method is only a slightly less favourable alternative to the PLS method when the errors are symmetric.
On the other hand, when the errors are distributed in an extremely skewed manner, the PCQR method could not out-perform the PQR method.
That is exactly the same situation as the fPC based methods when the CQR method is implemented.

\section{Real Data Analysis}

{\sl Real Data Analysis I: ADHD-200 fMRI Data. } We apply our proposed
method to a dataset on attention deficit hyperactivity disorder (ADHD)
from the ADHD-200 Sample Initiative Project. ADHD is the most commonly
diagnosed behavioral disorder of childhood, and can continue through
adolescence and adulthood. The symptoms include lack of attention,
hyperactivity, and impulsive behavior. The dataset we use is the
filtered preprocessed resting state data from New York University
(NYU) Child Study Center using the Anatomical Automatic Labeling (AAL)
\cite{tzourio2002automated} atlas. AAL contains 116 Regions of
Interests (ROI) fractionated into functional space using
nearest-neighbor interpolation. After cleaning the raw data that
failed in quality control or has missing data, we include 120
individuals in the analysis.

The response of interest is the ADHD index, Conners' parent rating scale-revised, long version (CPRS-LV), a continuous behavior score reflecting the severity of the ADHD disease. In the AAL atalas data,  the mean of the grey scale in each region is calculated for 172 equally spaced time points. We choose six parts of the brain which contain at least 4 ROIs, namely cerebelum, temporal, vermis, parietal, occipital, and frontal. The six functional predictors for each candidate part are computed by taking the average grey scale of the ROIs corresponding to each part, see Figure \ref{nc_fig1} (Left) for some selected subjects at cerebellum. The scalar covariates of primary interest include gender (female/male), age, handedness (continuous between -1 and 1, where -1 denotes totally left-handed and 1 denotes totally right-handed), diagnosis status (categorical with 3 levels: ADHD-combined, ADHD-inattentative and Control as baseline), medication status (yes/no), Verbal IQ, Performance IQ and Full4 IQ.  We build model to predict ADHD index adjusting these 9 scalar covariates (coded with dummy variables) using each of the six functional predictors. We consider the models for each individual functional covariates adjusting for the 9 scalar covariates. 

\begin{figure}[htbp]
 \centering
  \begin{minipage}[t]{.45\textwidth}
     \includegraphics[scale=0.33]{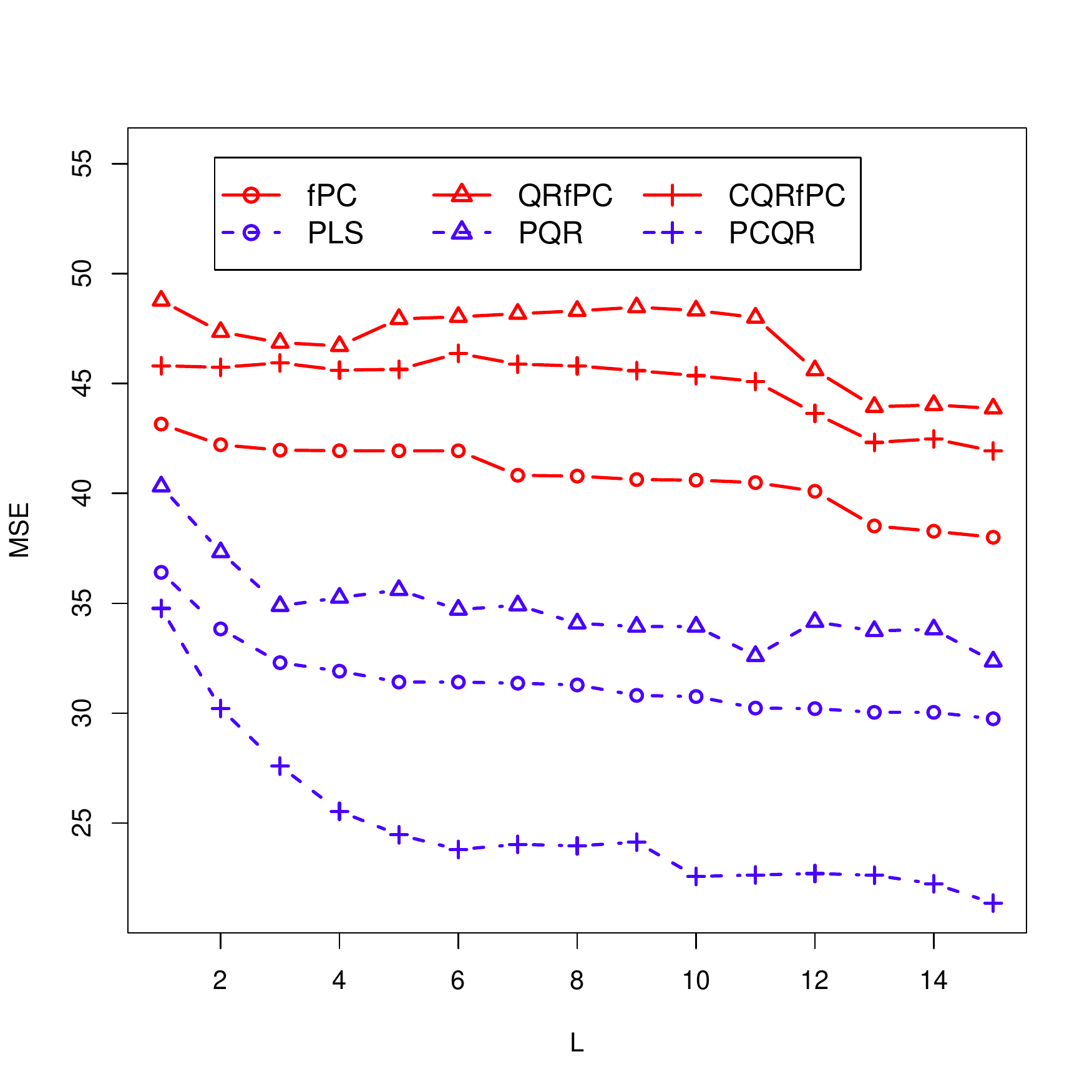}
 \end{minipage}
   \begin{minipage}[t]{.45\textwidth}
     \includegraphics[scale=0.33]{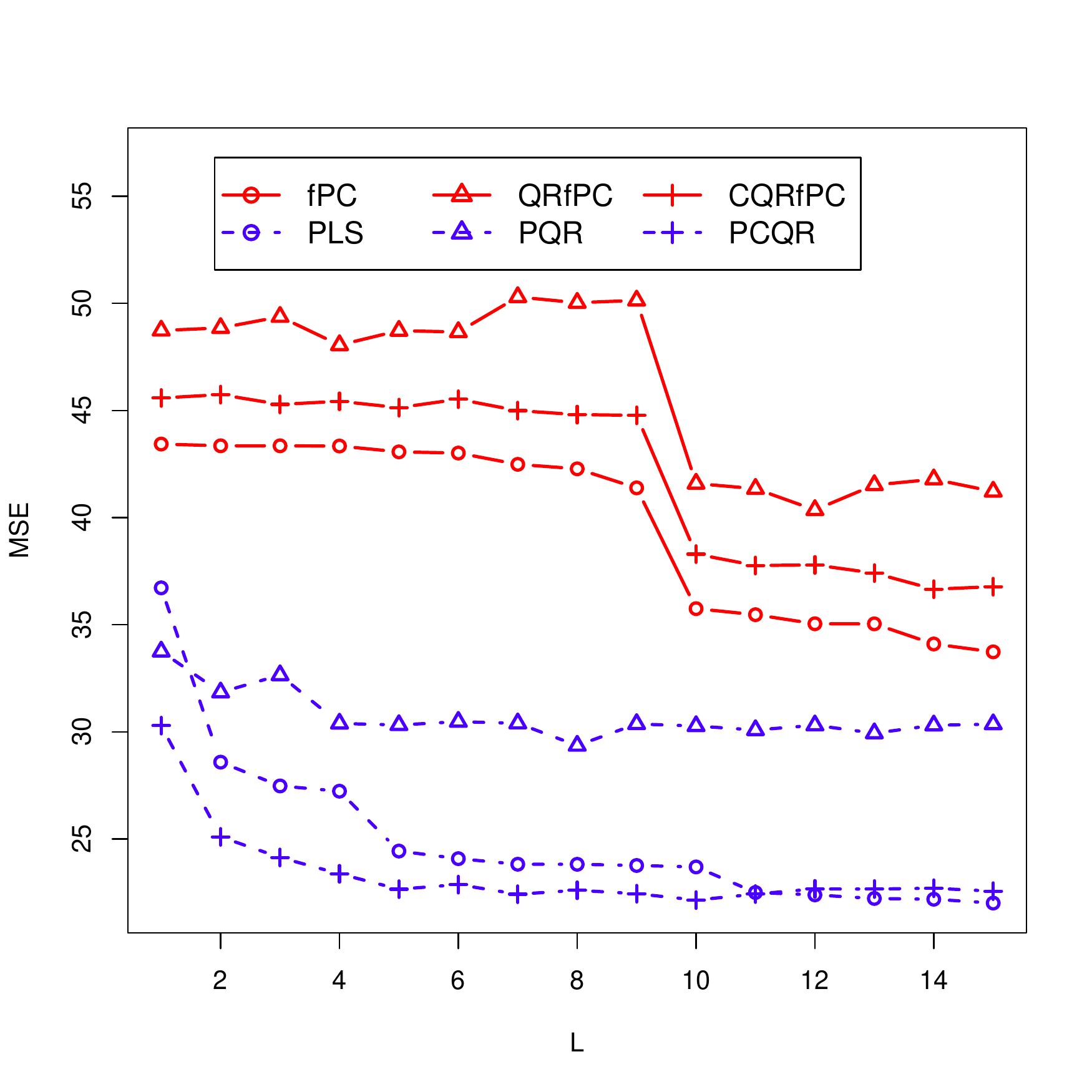}
 \end{minipage}
   \begin{minipage}[t]{.45\textwidth}
     \includegraphics[scale=0.33]{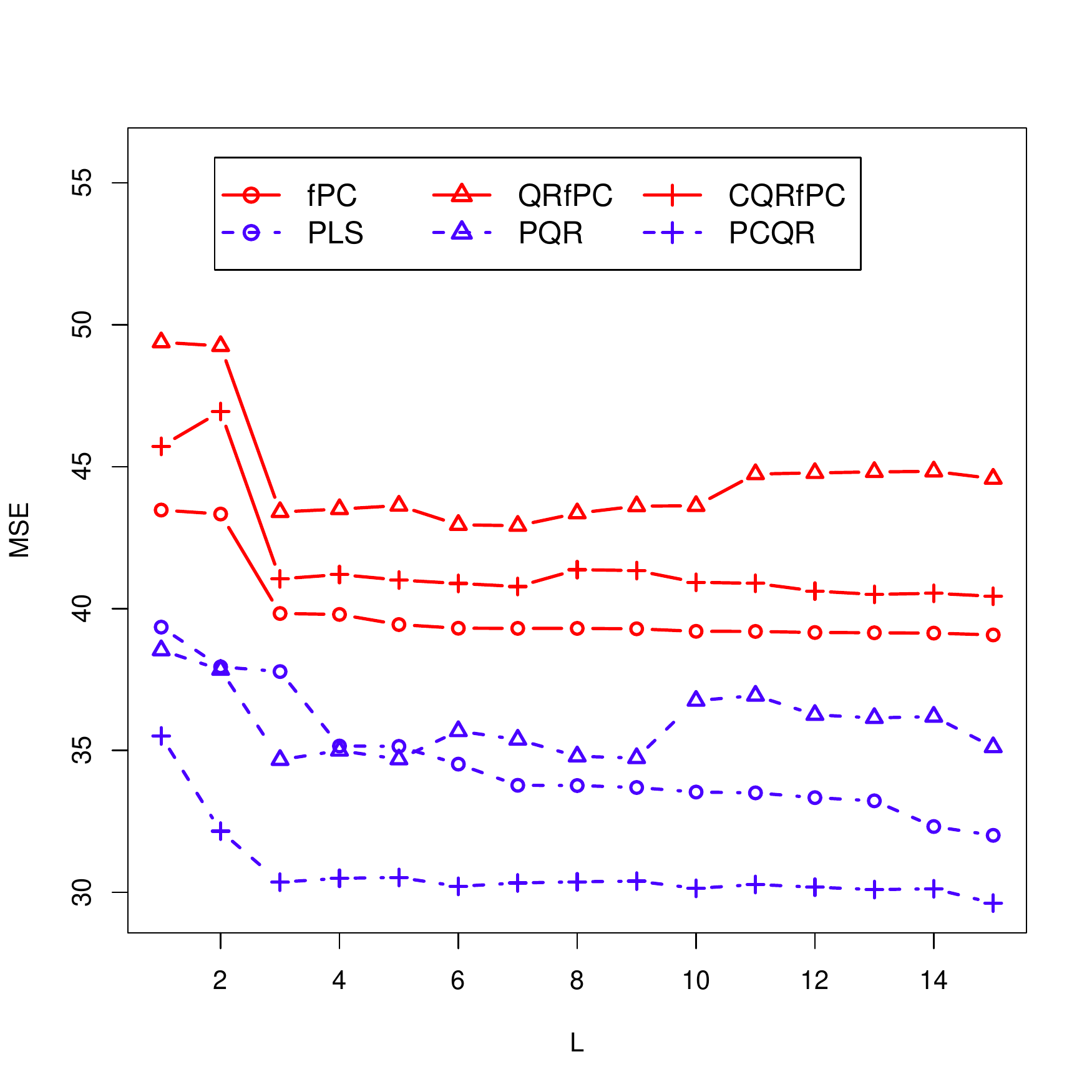}
 \end{minipage}
   \begin{minipage}[t]{.45\textwidth}
     \includegraphics[scale=0.33]{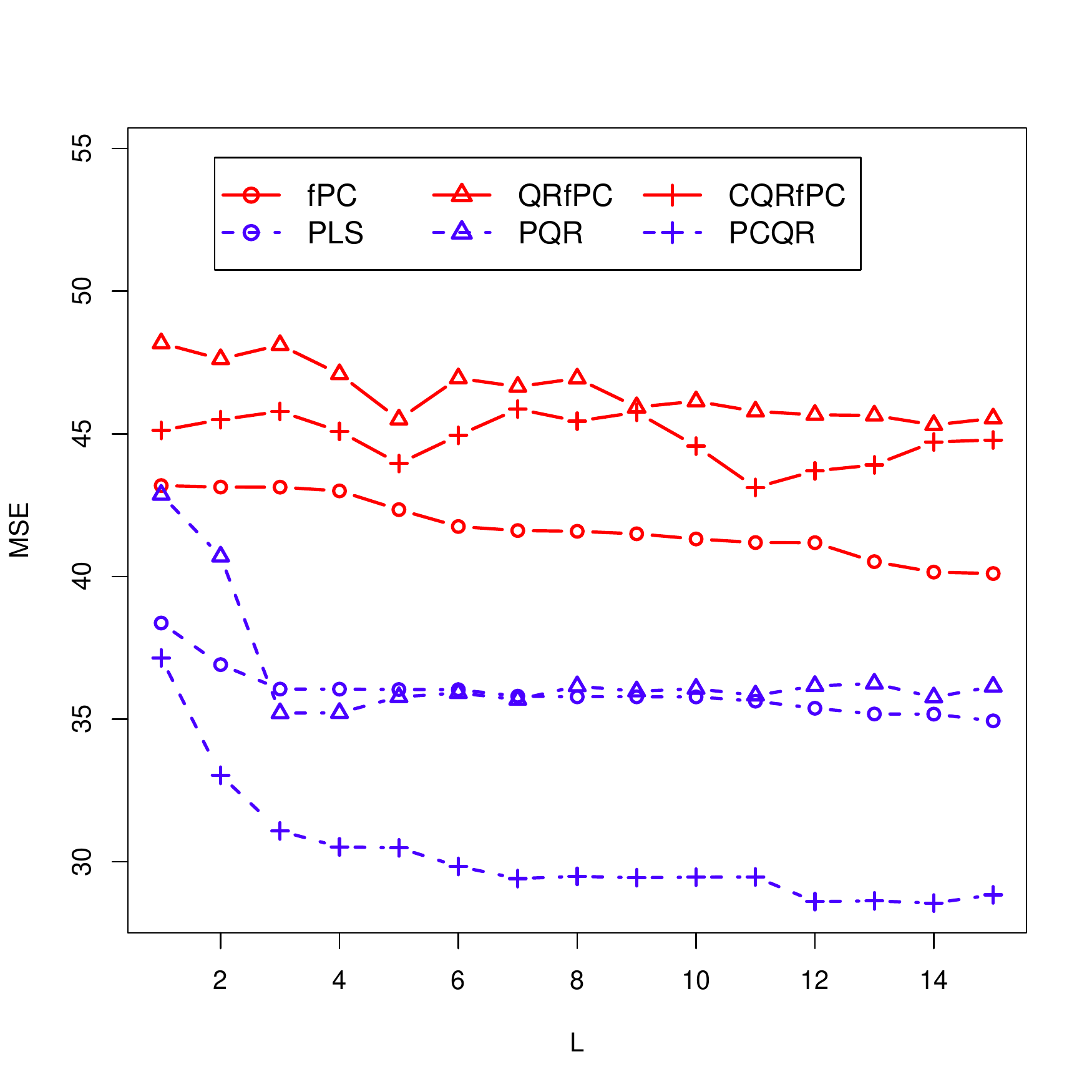}
 \end{minipage}
  \begin{minipage}[t]{.45\textwidth}
     \includegraphics[scale=0.33]{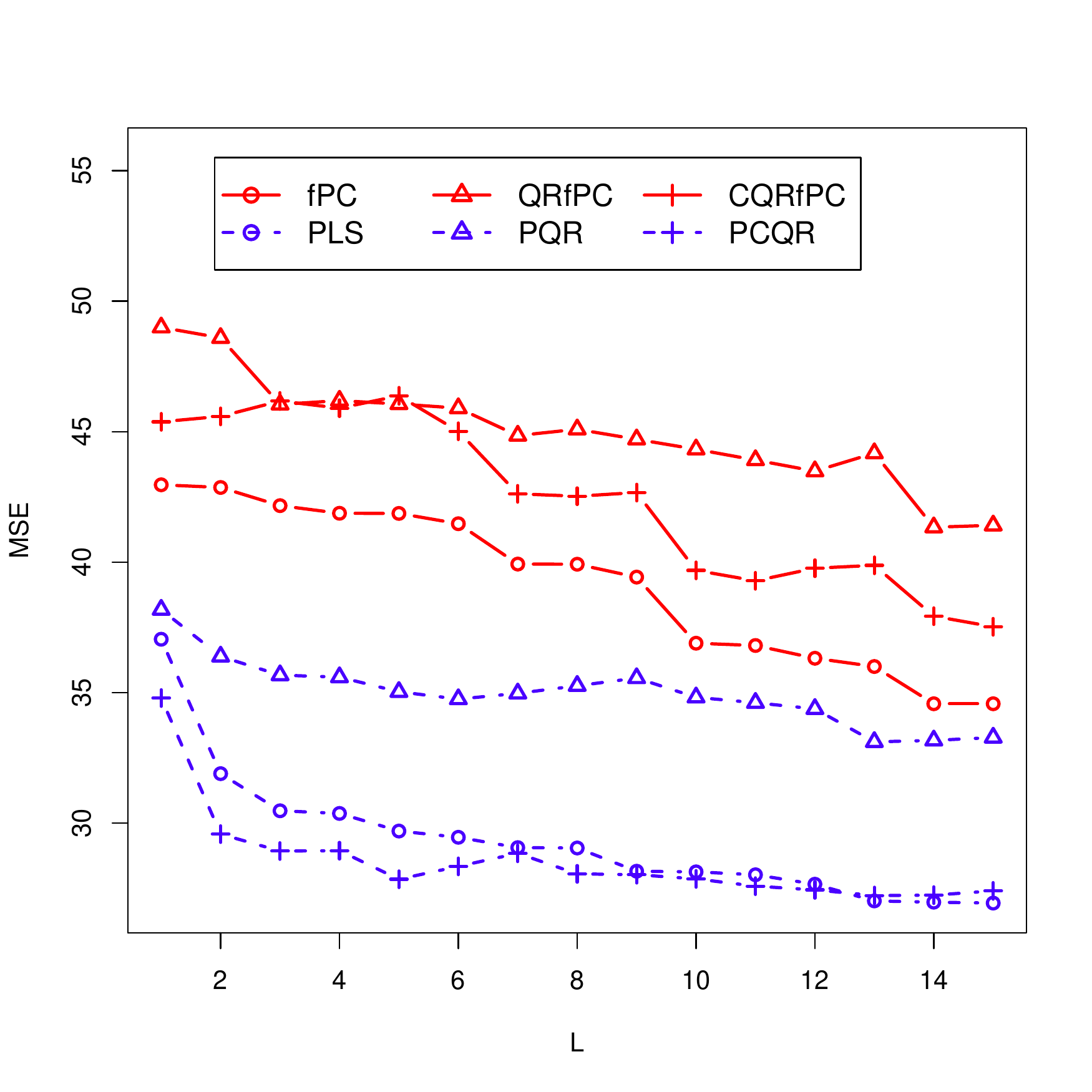}
 \end{minipage}
    \begin{minipage}[t]{.45\textwidth}
     \includegraphics[scale=0.33]{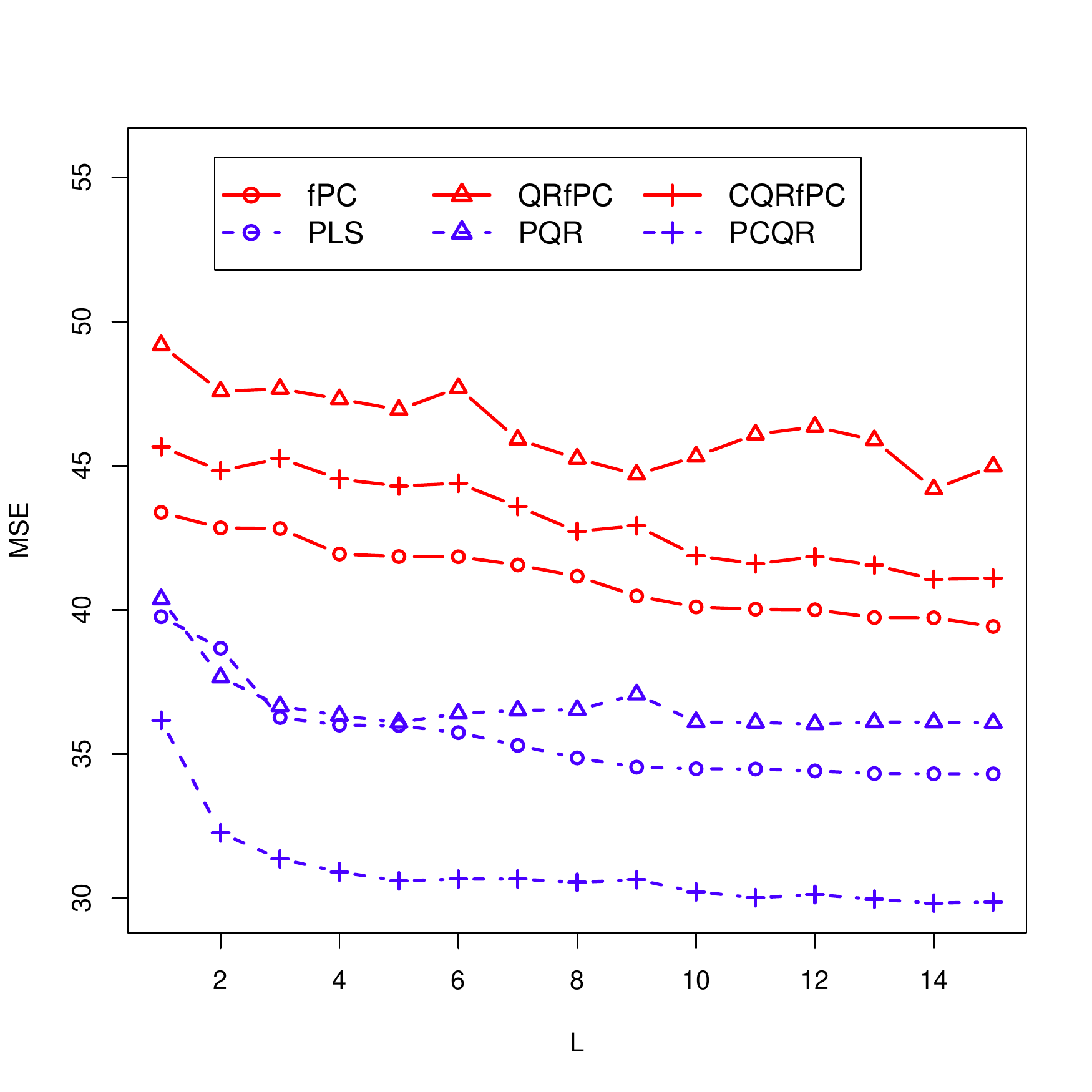}
 \end{minipage}
  \caption{Real Data Analysis I: ADHD-200 fMRI data, From up to down there are cerebelum, vermis, and occipital on the left panels and 
  temporal, parietal, and frontal on the right panel.}
  \label{adhdf6}
\end{figure}

Figure \ref{adhdf6} displays the changes of MSEs for all six methods,
with the increase of the number of cutoff level $L$ for different brain regions.
Here the quantile level $\tau$ is chosen to be fixed as $0.5$. As shown in the figure,  PLS, PQR and PCQR  methods perform much better than the fPC based methods while PCQR shows a significant superiority. In general, for each method only a few basis functions is capable of predicting the response well and additional basis functions 
do not decrease MSE much. This is more obvious for PLS, PQR and PCQR  methods as they consider information from the response while choose basis functions. 

{\sl Real Data Analysis II:  ADNI DTI Data. }  We use our model methods to analyze a real DTI data set with $n = 214$ subjects collected
from NIH Alzheimer's Disease Neuroimaging Initiative (ADNI) study. Data used in the preparation of this article were obtained from the Alzheimer's Disease Neuroimaging
Initiative (ADNI) database (adni.loni.ucla.edu). The ADNI was launched in 2003 by the National Institute on
Aging (NIA), the National Institute of Biomedical Imaging and Bioengineering (NIBIB), the Food and Drug
Administration (FDA), private pharmaceutical companies and non-profit organizations, as a \$60 million,
5-year public private partnership. The primary goal of ADNI has been to test whether serial magnetic
resonance imaging (MRI), positron emission tomography (PET), other biological markers, and clinical and
neuropsychological assessment can be combined to measure the progression of mild cognitive impairment
(MCI) and early Alzheimer's disease (AD). Determination of sensitive and specific markers of very early
AD progression is intended to aid researchers and clinicians to develop new treatments and monitor their
effectiveness, as well as lessen the time and cost of clinical trials. The Principal Investigator of this initiative
is Michael W. Weiner, MD, VA Medical Center and University of California, San Francisco. ADNI is the
result of efforts of many coinvestigators from a broad range of academic institutions and private corporations,
and subjects have been recruited from over 50 sites across the U.S. and Canada. The initial goal of ADNI was
to recruit 800 subjects but ADNI has been followed by ADNI-GO and ADNI-2. To date these three protocols
have recruited over 1500 adults, ages 55 to 90, to participate in the research, consisting of cognitively normal
older individuals, people with early or late MCI, and people with early AD. The follow up duration of
each group is specified in the protocols for ADNI-1, ADNI-2 and ADNI-GO. Subjects originally recruited
for ADNI-1 and ADNI-GO had the option to be followed in ADNI-2. For up-to-date information, see {\sl www.adni-info.org}.
The significance level is an ongoing public-private partnership to test whether genetic, structural and functional
neuroimaging, and clinical data can be integrated to assess the progression of mild cognitive
impairment (MCI) and early Alzheimer's disease (AD). The structural brain MRI data and
corresponding clinical and genetic data from baseline and follow-up were downloaded from
the ADNI publicly available database ({\sl https://ida/loni/usc/edu}).

\begin{figure}[!ht]
\includegraphics[height = .36\textheight,width=.50\textwidth]{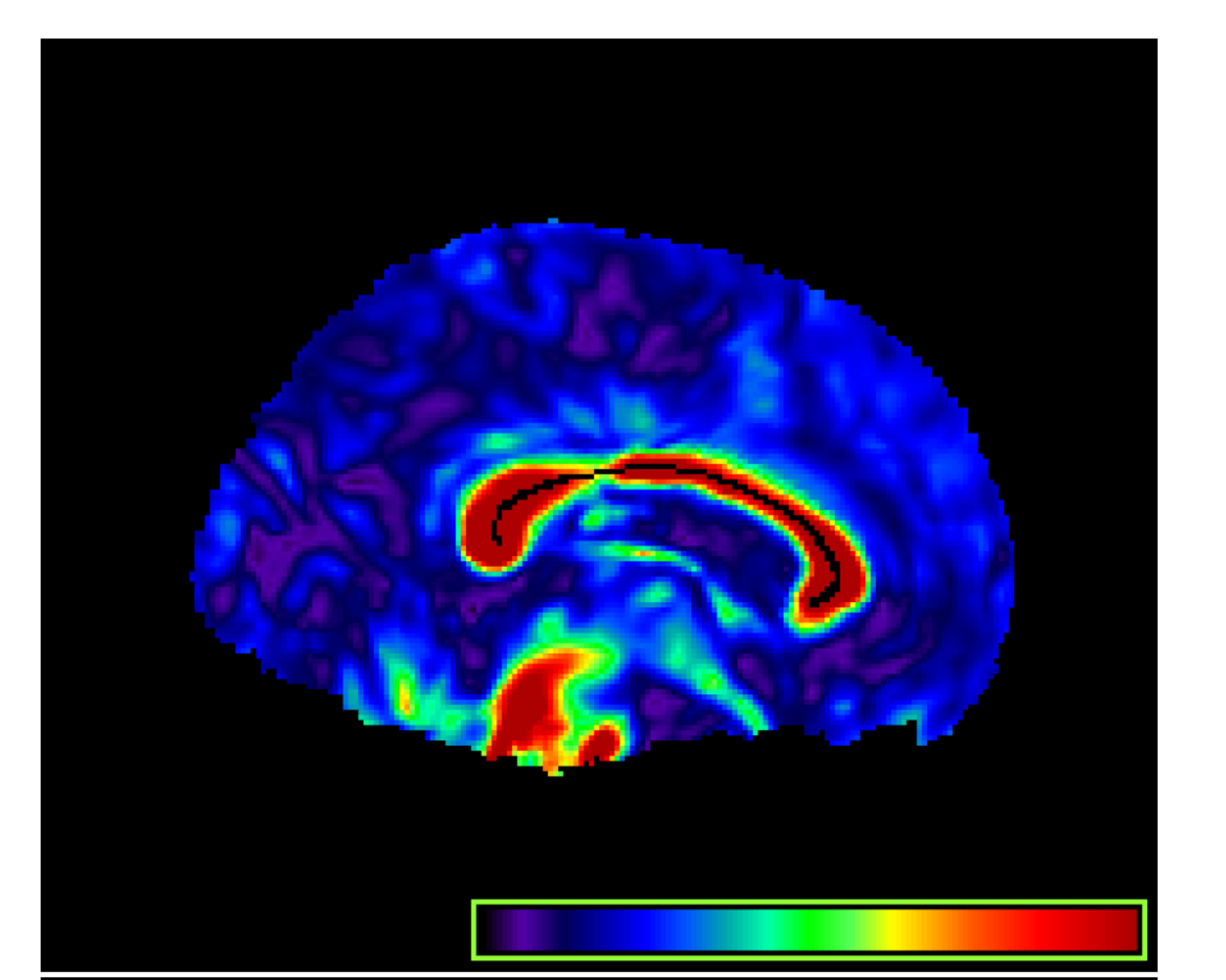}%
\includegraphics[height = .36\textheight,width=.50\textwidth]{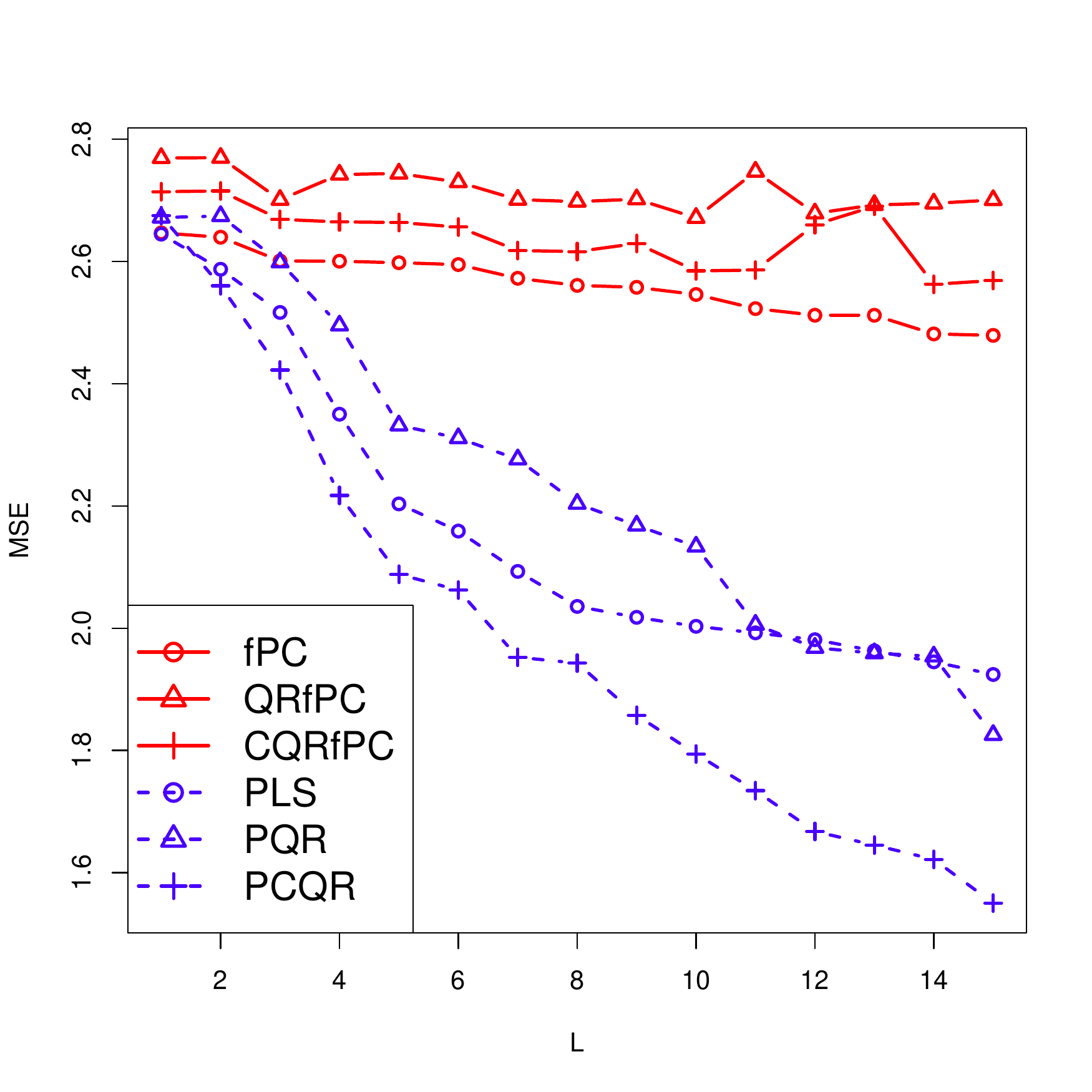}
\caption{ Real Data Analysis II:  (Left)  The midsagittal corpus callosum (CC) skeleton overlaid with fractional anisotropy (FA) from one randomly selected subject and (Right) the MSE of mini-mental state examination (MMSE) at different cut-off levels.} \label{adnif7}
\end{figure}

The DTI data were processed by two key steps including a weighted least squares estimation
method Basser et al. \cite{basser1994estimation}; Zhu et al. \cite{zhu2007statistical} to construct the diffusion tensors
and a FSL TBSS pipeline Smith et al. \cite{smith2006tract} to register DTIs from multiple subjects to
create a mean image and a mean skeleton. Speciffically, maps of fractional anisotropy (FA)
were computed for all subjects from the DTI after eddy current correction and automatic
brain extraction using FMRIB software library. FA maps were then fed into the TBSS tool,
which is also part of the FSL. In the TBSS analysis, the FA data of all the subjects were
aligned into a common space by non-linear registration and the mean FA image were created
and thinned to obtain a mean FA skeleton, which represents the centers of all WM tracts
common to the group. Subsequently, each subjects aligned FA data were projected onto this
skeleton. We focus on the midsagittal corpus callosum skeleton and associated FA curves from all
subjects, see Figure \ref{nc_fig1} (Right) for some selected subjects. The corpus callosum (CC) is the largest fiber 
tract in the human brain and is a topographically organized structure, see Figure \ref{adnif7} (Left). It is responsible for much of
the communication between the two hemispheres and connects homologous areas in the two
cerebral hemispheres. It is important in the transfer of visual, motoric, somatosensory, and auditory information.

We are interested in predicting mini-mental state examination (MMSE) scores, one of the most widely used screening tests, which are used to provide brief, objective measures of cognitive functioning for almost fifty years.
The MMSE scores has been seen as a reliable and valid clinical measure quantitatively assessing the severity of cognitive impairment.
It was believed that the MMSE scores to be affected by demographic features
such as age, education and cultural background, but not gender \cite{tombaugh1992mini}. After quality control and excluding the missing data, 
we include 200 subjects from the total 217 subjects. 
The functional covariate is fractional anisotropy (FA) values along the  corpus callosum (CC) fiber tract with 83 equally spaced grid points,
which can be treated as a function of arc-length. The scale covariates are the
gender variable (coded by a dummy variable indicating for male), the age of the subject (years), the education level (years), an indicator for Alzheimer's disease
(AD) status (19.6\%) and an indicator for mild cognitive impairment (MCI) status (55.1\%), and genotypes for apolipoprotein E $\epsilon$-$4$ (coded by three indicator variables for four levels). 

The MSEs are shown in figure \ref{adnif7}. In general, PLS, PQR and PCQR  methods present consistently better than fPC based methods while
PCQR outperforms PQR and PCQR  methods.
The phenomenon has been observed from the previous read data analysis, which indicates that for brain imaging data 
PCQR method has a improved prediction accuracy compared with PQR and PCQR  methods. With the number of functional basis increases, 
the MSEs do not decreases much for fPC based methods while constantly decrease for PLS, PQR and PCQR  methods.  This indicates 
that the fPC basis is not suitable to do prediction though they may account a large portion of the variations of functional covariates. The PLS, PQR and PCQR  methods 
is capable of explaining a large percentage variation of the response and reducing the MSEs by proving appropriate basis functions. Our proposed methods show 
great superiority to the fPC based methods and the PLS methods and provide a powerful tool to do prediction in practice.

\section{Discussion}

In this paper, we first define the concept partial quantile covariance (PQC) to measure the contribution of one covariate to the response. 
We then propose the partial functional linear quantile regression method to use partial quantile regression (PQR) to extract PQR basis to 
effectively predict the response. This is motivated by the success of the partial least square (PLS) basis in functional linear regression model.  
The key idea is to use both information from the functional covariates and the response and therefore both PQR basis and PLS basis can be treated 
as supervised learning while fPC based methods are semi-supervised learning as they only use information from the functional covariates. The 
algorithm SIMPQR we developed is analogue to that of SIMPLS. We extend PQC to partial composite quartile covariance (PCQC) and 
propose the PCQR basis and its SIMPCQR algorithm under the homoscedasticity condition.  

The simulations show that PLS, PQR and PCQR in general perform better than the fPC based methods. However, PQR method is more robust against skewness of error distribution while the PLS and  PCQR methods act similarly to each other and perform better than PQR method when the error distribution is symmetric.
This advantage from PQR method can be explained by the general nature of quantile method which obtains its robustness by sacrificing certain efficiency.
By assuming homoscedasticity, the PCQR method acts similarly to the PLS method when the error distributions are symmetric
but retains its robustness when the error distributions are extremely skewed. 

Our proposed methods, PQR and PCQR methods, significantly outperform other methods, especially those fPC based methods in both ADHD-200 fMRI 
data analysis and ADNI DTI data analysis.  In ADHD-200 fMRI data analysis, our methods are capable of reducing much more MSEs by using only a few basis while 
fPC based method are not even by adding more basis. In 
ADNI DTI data analysis, both PQR and PCQR methods reduce significant amount MSEs with more and more basis. On the other hand, fPC based 
methods perform poor even with more basis.  Overall in the two neuroimaging data analysis, PCQR 
performs slightly better than PQR though.

The consistency of the PLS methods was proved by Delaigle and Hall \cite{delaigle2012methodology} where they found an equivalent space with explicit expressed basis functions 
to the PLS basis space. For PQR and PCQR methods, it is difficult to find such equivalent space and therefore their consistency may not be easy to show. The difficulty 
of the problem lies on the iterative nature of PQR and PCQR methods where basis is sequentially extracted.  One way to overcome that is to find preselected number of basis 
simultaneously \cite{zhou2013tensor}.  Another direction is to impose certain structure 
on the selected basis, for example, sparsity and smoothness in PLS methods \cite{reiss2007functional}. This can be done for simultaneous basis selection as well \cite{zhou2014regularized}. 

In both simulation studies and real data analysis, only univariate functional covariate case is considered.
However, the extension of PQR and PCQR methods to multivariate functional covariates is straightforward. 
The computation becomes more complex and intensive due to the iterative basis extraction nature.
Such complexity is expected to be significantly reduced 
by applying simultaneous basis selection or imposing certain structure on the selected basis. Further details are out of the scope of this manuscript
 and will be pursuit in the future research.

\section*{Acknowledgments}
Dengdeng Yu and Dr. Linglong Kong's  research were supported by the startup from the University of Alberta and grants from the
Natural Sciences and Engineering Research Council of Canada. Dr. Ivan Mizera's research was supported by grants from the
Natural Sciences and Engineering Research Council of Canada. Dr. Linglong Kong also wants to thank the support of the Program on Low-dimensional 
Structure in High-dimensional Systems (LDHD) at the Statistical and Applied Mathematical Sciences Institute (SAMSI) during his visit in 2014. 

Part of data collection and sharing for this project was funded by the Alzheimer's Disease 
Neuroimaging Initiative (ADNI) (National Institutes of Health Grant U01 AG024904). ADNI
is funded by the National Institute on Aging, the National Institute of Biomedical Imaging
and Bioengineering, and through generous contributions from the following: Alzheimer's
Association; Alzheimer's Drug Discovery Foundation; BioClinica, Inc.; Biogen Idec Inc.;
Bristol-Myers Squibb Company; Eisai Inc.; Elan Pharmaceuticals, Inc.; Eli Lilly and Company;
F. Hoffmann-La Roche Ltd and its affilated company Genentech, Inc.; GE Healthcare;
Innogenetics, N.V.; IXICO Ltd.; Janssen Alzheimer Immunotherapy Research \& Development,
LLC.; Johnson \& Johnson Pharmaceutical Research \& Development LLC.; Medpace,
Inc.; Merck \& Co., Inc.; Meso Scale Diagnostics, LLC.; NeuroRx Research; Novartis Pharmaceuticals
Corporation; Pfizer Inc.; Piramal Imaging; Servier; Synarc Inc.; and Takeda
Pharmaceutical Company. The Canadian Institutes of Health Research is providing funds
to support ADNI clinical sites in Canada. Private sector contributions are facilitated by the
Foundation for the National Institutes of Health ({\sl www.fnih.org}). The grantee organization
is the Northern California Institute for Research and Education, and the study is coordinated by the 
Alzheimer's Disease Cooperative Study at the University of California, San
Diego. ADNI data are disseminated by the Laboratory for Neuro Imaging at the University
of California, Los Angeles. This research was also supported by NIH grants P30 AG010129
and K01 AG030514.

\label{ns_re}




\section*{References}
\bibliographystyle{elsarticle-harv} 
 \bibliography{neuro_lk.bib}


%\begin{thebibliography}{00}
%%% Text of bibliographic item
%
%\bibitem{}
%
%\end{thebibliography}

\newpage

\begin{table}[htbp]
\centering
\small
\addtolength{\tabcolsep}{-2pt}
\begin{tabular}{|| c| c|c ||   l l l   |   l l l   |  l l l  |   l l l  |}
\hline
  \multicolumn{3}{||c||}{}  & \multicolumn{3}{|c|}{fPC}      & \multicolumn{3}{|c|}{QRfPC} & \multicolumn{3}{|c|}{CQRfPC}\\
 \hline
 Error & L &$n$ &$\rm{Bias}^2$ & Var & MISE& $\rm{Bias}^2$ & Var & MISE & $\rm{Bias}^2$ & Var & MISE  \\
\hline
\multirow{9}{*}{Gaussian} & \multirow{3}{*}{1} & 100 & 3.63 & 0.07* & 3.70 & 3.63 & 0.09 & 3.72 & 3.63 & 0.07* & 3.71 \\ 
    & & 200 & 3.64 & 0.03* & 3.67 & 3.63 & 0.04 & 3.68 & 3.63 & 0.04 & 3.67 \\ 
    & & 500 & 3.69 & 0.02* & 3.70 & 3.68 & 0.02* & 3.71 & 3.69 & 0.02* & 3.70 \\ 
    \cline{2-12}
   &  \multirow{3}{*}{2} & 100 & 0.78 & 0.36* & 1.14* & 0.77 & 0.39 & 1.16 & 0.78 & 0.36* & 1.14* \\ 
   &  & 200 & 0.86 & 0.17* & 1.03 & 0.86 & 0.18 & 1.04 & 0.86 & 0.17* & 1.03 \\ 
   &  & 500 & 0.86 & 0.09* & 0.95 & 0.86 & 0.10 & 0.96 & 0.86 & 0.09* & 0.95 \\
   \cline{2-12}
   &  \multirow{3}{*}{3} & 100 & 0.32 & 0.34* & 0.67* & 0.33 & 0.38 & 0.70 & 0.32 & 0.35 & 0.67* \\ 
   &  & 200 & 0.28 & 0.19* & 0.47* & 0.28 & 0.22 & 0.50 & 0.28 & 0.20 & 0.48 \\ 
   &  & 500 & 0.29 & 0.08* & 0.38 & 0.29 & 0.09 & 0.38 & 0.29 & 0.08* & 0.38 \\ 
   \hline
   \multirow{9}{*}{Cauchy} & \multirow{3}{*}{1} & 100 & 7.32 & $>$100 & $>$100 & 3.65 & 0.12 & 3.76 & 3.66 & 0.09* & 3.75 \\ 
   &  & 200 & 4.82 & 54.42 & 59.24 & 3.63 & 0.10 & 3.73 & 3.64 & 0.08* & 3.72 \\ 
   &  & 500 & 43.78 & $>$100 & $>$100 & 3.71 & 0.03 & 3.74 & 3.72 & 0.02* & 3.74 \\ 
   \cline{2-12}
   & \multirow{3}{*}{2}  & 100 & 7.56 & $>$100 & $>$100 & 0.77 & 0.40* & 1.17* & 0.76 & 0.42 & 1.18 \\ 
   &  & 200 & 2.14 & $>$100 & $>$100 & 0.78 & 0.21* & 0.99 & 0.77 & 0.22 & 1.00 \\ 
   &  & 500 & $>$100 & $>$100 & $>$100 & 0.81 & 0.10* & 0.91 & 0.81 & 0.10* & 0.91 \\ 
   \cline{2-12}
   & \multirow{3}{*}{3}  & 100 & 6.03 & $>$100 & $>$100 & 0.28 & 0.48* & 0.76* & 0.27 & 0.53 & 0.80 \\ 
   &  & 200 & 3.67 & $>$100 & $>$100 & 0.31 & 0.24* & 0.55* & 0.31 & 0.27 & 0.58 \\ 
   &  & 500 & $>$100 & $>$100 & $>$100 & 0.31 & 0.11* & 0.41* & 0.31 & 0.12 & 0.42 \\ 
     \hline
      \hline
  \multicolumn{3}{||c||}{}  & \multicolumn{3}{|c|}{PLS}      & \multicolumn{3}{|c|}{PQR} & \multicolumn{3}{|c|}{PCQR}\\
 \hline
 Error & L &$n$ &$\rm{Bias}^2$ & Var & MISE& $\rm{Bias}^2$ & Var & MISE & $\rm{Bias}^2$ & Var & MISE  \\
 \hline
\multirow{9}{*}{Gaussian} & \multirow{3}{*}{1} & 100 & 0.54 & 0.82 & 1.36* & 0.50* & 0.91 & 1.41 & 0.52 & 0.84 & 1.36* \\ 
   &  & 200 & 0.63 & 0.20 & 0.83 & 0.57* & 0.26 & 0.83 & 0.60 & 0.21 & 0.81* \\ 
   &  & 500 & 0.59 & 0.07 & 0.66 & 0.52* & 0.10 & 0.62* & 0.56 & 0.08 & 0.64 \\ 
    \cline{2-12}
   &  \multirow{3}{*}{2} & 100 & 0.11* & 1.07 & 1.18 & 0.15 & 1.18 & 1.33 & 0.12 & 1.07 & 1.19 \\ 
   &  & 200 & 0.12* & 0.29 & 0.41* & 0.16 & 0.36 & 0.52 & 0.13 & 0.29 & 0.43 \\ 
   &  & 500 & 0.11* & 0.10 & 0.21* & 0.14 & 0.13 & 0.27 & 0.12 & 0.10 & 0.23 \\ 
    \cline{2-12}
   &  \multirow{3}{*}{3} & 100 & 0.08 & 2.58 & 2.66 & 0.07 & 2.28 & 2.36 & 0.06* & 2.65 & 2.71 \\ 
   &  & 200 & 0.04* & 0.87 & 0.91 & 0.04* & 1.10 & 1.13 & 0.04* & 0.96 & 0.99 \\ 
   &  & 500 & 0.02* & 0.26 & 0.28* & 0.02* & 0.43 & 0.45 & 0.02* & 0.30 & 0.32 \\ 
  \hline
\multirow{9}{*}{Cauchy} & \multirow{3}{*}{1} & 100 & 71.08 & $>$100 & $>$100 & 0.49 & 1.23 & 1.72* & 0.48* & 1.39 & 1.87 \\ 
   &  & 200 & 41.24 & $>$100 & $>$100 & 0.48 & 0.43 & 0.91* & 0.47* & 0.47 & 0.94 \\ 
   &  & 500 & $>$100 & $>$100 & $>$100 & 0.46* & 0.16 & 0.62* & 0.48 & 0.17 & 0.64 \\ 
    \cline{2-12}
   &  \multirow{3}{*}{2} & 100 & $>$100 & $>$100 & $>$100 & 0.16 & 1.96 & 2.12 & 0.12* & 3.01 & 3.13 \\ 
   &  & 200 & $>$100 & $>$100 & $>$100 & 0.15 & 0.67 & 0.82* & 0.12* & 0.98 & 1.09 \\ 
   &  & 500 & $>$100 & $>$100 & $>$100 & 0.14 & 0.23 & 0.37* & 0.11* & 0.29 & 0.41 \\ 
    \cline{2-12}
   &  \multirow{3}{*}{3} & 100 & $>$100 & $>$100 & $>$100 & 0.13* & 5.84 & 5.97 & 0.20 & 14.28 & 14.48 \\ 
   &  & 200 & $>$100 & $>$100 & $>$100 & 0.06* & 2.59 & 2.65 & 0.11 & 5.45 & 5.56 \\ 
   &  & 500 & $>$100 & $>$100 & $>$100 & 0.02* & 1.01 & 1.02 & 0.04 & 1.67 & 1.71 \\ 
\hline
\end{tabular}
\caption{Simulation I: $*$ flags the minimum values of the six methods in each measurement category of $\rm{Bias}^2$, Var and MISE.}
\label{sim1t1}
\end{table}

\end{document}